\documentclass[acmsmall]{acmart}

\usepackage{subcaption}

\usepackage{booktabs}
\usepackage{adjustbox}

\usepackage{hhline} 
\usepackage{longtable}
\usepackage{multirow}
\usepackage{lscape} 
\usepackage{geometry} 
\usepackage{setspace} 
\usepackage[normalem]{ulem} 

\AtBeginDocument{%
  }

\setcopyright{cc}
\setcctype{by}
\acmJournal{PACMHCI}
\acmYear{2026}
\acmVolume{10}
\acmNumber{6}
\acmArticle{CSCW048}
\acmMonth{10}
\acmDOI{10.1145/3816896}

\begin{document}

\title{AI in the Workplace: The Impact of AI on Perceived Job Decency and Meaningfulness}


\author{Kuntal Ghosh}
\email{kuntal.ghosh@uni-siegen.de}
\affiliation{%
  \institution{University of Siegen}
  \streetaddress{Kohlbettstraße 15}
  \city{Siegen}
  \postcode{57072}
  \country{Germany}
}

\author{Marc Hassenzahl}
\email{marc.hassenzahl@uni-siegen.de}
\affiliation{%
  \institution{University of Siegen}
  \streetaddress{Kohlbettstraße 15}
  \city{Siegen}
  \postcode{57072}
  \country{Germany}
}

\author{Shadan Sadeghian}
\email{shadan.sadeghian@uni-siegen.de}
\affiliation{%
  \institution{University of Siegen}
  \streetaddress{Kohlbettstraße 15}
  \city{Siegen}
  \postcode{57072}
  \country{Germany}
}

\renewcommand{\shortauthors}{Ghosh et al.}

\begin{abstract}
    The proliferation of Artificial Intelligence (AI) in workplaces is transforming how we work.
    While existing research on human-AI collaboration at work often prioritizes performance, less is known about their experiential outcomes. 
    Through interviews with 24 employees across Information Technology (IT), service-based, and healthcare sectors, this paper examines AI's impact on job satisfaction via perceptions of job decency and meaningfulness, now and in the future.
    Our results reveal that the anticipated impact of AI on overall job satisfaction varies with the occupational domain, with differing perceptions of its underlying decency and meaningfulness. For instance, IT and healthcare anticipate increased satisfaction with decency aspects like working hours but decreased satisfaction with meaningfulness aspects like social image due to misconceptions about AI handling most of their tasks. Conversely, service workers foresee no improvement in their working hours but a higher social standing due to the perceived status boost associated with working with AI.
\end{abstract}



\received{May 13, 2025}
\received[revised]{January 13, 2026}
\received[accepted]{March 17, 2026}

\maketitle

\section{Introduction}
With more than one-third of the human lifespan dedicated to work, it is evident that job satisfaction plays an integral role in shaping our life experiences \cite{ahmad2013paradigms}. Besides providing financial stability, work is a crucial source of meaning and satisfaction \cite{terkel1974working}. Thus having a job that serves a higher purpose leads to higher job satisfaction \cite{sparks2001explaining, lips2009discriminating, laschke2020positive}.
Previous research highlights two fundamental dimensions of contemporary work - job decency and job meaningfulness \cite{blustein2023understanding}.
Decent work outlines the minimum workplace criteria that ought to be associated with work, while meaningful work represents the desire to find significance and purpose in one's professional endeavors. Both of these sources are multi-dimensional constructs that can be influenced by various factors. For example, while the perception of job decency is influenced by factors such as working hours, work-life balance, job security, growth opportunities, remuneration, and working conditions \cite{duffy2016psychology}, job meaningfulness is shaped by the extent to which employees feel accomplished in their tasks, encounter stimulating challenges that foster meaningful contributions, maintain positive relationships with colleagues, and take pride in their affiliation with their organizations \cite{smids2020robots, baldauf2021automation}.
Building on this, we understand job satisfaction as arising, in part, from workers’ experiences of decency and meaningfulness in their roles.

Technology has long played a pivotal role in the evolution of work, evolving from primitive tools to contemporary computing devices. In the last years, the advent of artificial intelligence (AI) has made it an essential element for consideration in many work domains \cite{dwivedi2021artificial, marquis2024proliferation} and will lead to inevitable transitions in work practices. These transitions can consequently influence workers' job satisfaction, specifically in domains where the labor was solely conducted by human workers such as call center support, or project resource management \cite{miller2018ai,  daugherty2018human+}. However, research involving human-AI collaboration (HAIC) tends to prioritize performance-related factors, such as efficiency and effectiveness, over the experiential aspects of work \cite{sadeghian2022artificial}. While this may be beneficial in the short term, it might lead to workers' disengagement and dissatisfaction in the long term \cite{parasuraman2008situation}.
There are only a few research works that address job satisfaction in collaboration with AI (e.g., \cite{walliser2019team, nyholm2020can, mitchell2021automated, sadeghian2024soul, waardenburg2024human}). However, none of these have looked into decency and meaningfulness as concurrent determining elements of job satisfaction and how they vary across different work domains.
Thus, it remains unclear how the introduction of AI in work practices changes the perception of job satisfaction, particularly regarding, decency and meaningfulness as two fundamental determinants. Therefore, in this paper, we attempt to answer the research question: \textbf{"How will the introduction of AI in workplaces change the perception of job decency and meaningfulness across different occupational domains?"}

We make the following contribution to CSCW research:
\begin{enumerate}
    \item A qualitative study using anticipatory ethnography and semi-structured interviews with 24 professionals from IT, service-based, and healthcare domains.
    \item A synthesized framework of job satisfaction grounded in existing literature, delineating job decency and job meaningfulness as key components, each with their underlying attributes, thereby offering a useful framework for future studies.
    \item Empirical insights into the professionals' perceptions of job decency and meaningfulness, both in the present and in an anticipated future of work with AI.
    \item Design recommendations for developing AI-driven systems aimed at supporting decent and meaningful work, emphasizing domain-specific values and worker expectations, diverging from "one-size-fits-all" approaches.
\end{enumerate}

The ensuing sections start with a summary of related work on decent and meaningful work, followed by job satisfaction in collaboration with AI. We then present our research method and subsequent results. Finally, we discuss our findings and conclude with design implications and a reflection on future work.

\section{Related Work}
\subsection{Job Satisfaction: Decency and Meaningfulness}
Work is an important part of daily life.
Drawing on Duffy et al. \cite{duffy2016psychology} and Allan et al. \cite{allan2020decent}, we understand it as fulfilling three elementary human needs: 
(1) Survival and power: the need for food and shelter, and access to opportunity structures like the education system, labor market, healthcare system, or technological access; 
(2) Social connections: the need to both feel, and be part of a community; and 
(3) Self-determination: the need for autonomy, and feeling of competence and empowerment.
Furthermore, work that provides a greater sense of purpose increases job satisfaction and consequently, the perception of meaning in life \cite{sparks2001explaining, lips2009discriminating, laschke2020positive}. Job satisfaction refers to the positive emotions a person feels when they find their job fulfilling \cite{locke1969job}. It leads to increased engagement and organizational commitment, enhanced employee performance, and lower attrition \cite{henne1985job, zopiatis2014job, kong2018job, dorta2021effects}. It can affect individuals positively in terms of mental health, and satisfying relationships \cite{kim2020thriving}. Existing literature underscores \textit{job decency} and\textit{ job meaningfulness} as determinants of job satisfaction \cite{blustein2023understanding}.

Decent work outlines the minimum workplace criteria that ought to be associated with work, which is synonymous with the four pillars of the International Labour Organization's (ILO) decent work agenda -  employment creation, workplace rights, social protection, and social dialogue \cite{office2014global}. Thus, the central theme of decent work is meeting basic human needs at the workplace. 
Aspects such as working hours, work-life balance, job security, growth opportunities, remuneration, and working conditions are entwined with ILO's decent work agenda and contribute to the overall concept of job decency \cite{duffy2016psychology}. Other factors are harmony between work life and personal life that alleviates stress and discomfort \cite{greenhaus2011work, allen2012work, greenhaus2014contemporary}, job security which prevents the negative effects of unemployment on social integration and mental health \cite{paul2009unemployment, wanberg2012individual, gowan2014moving}, growth opportunities which increases career management and hire-ability, and fosters a sense of autonomy among employees \cite{baruch2006career, sullivan2009advances}, fair remuneration that creates a sense of justice, motivates individuals in their jobs and aids in realizing an appropriate standard of living \cite{dulebohn2007compensation, judge2010relationship}. Thus, decent work involves opportunities to do productive work in a safe working environment that generates fair remuneration, job stability, social protection, social dialogue, social integration, and equality in terms of opportunities and conduct \cite{pereira2019empirical}.

Existing literature has provided various definitions for meaningful work. While some researchers have argued that meaningful work carries a positive connotation for individuals \cite{rosso2010meaning}, others mention personal growth and impact on broader welfare as indicators of meaningfulness \cite{steger2012measuring}. Various perspectives, such as answering one's calling \cite{dik2009calling, lysova2019meaningful}, and comprehending one's reason for existence \cite{pratt2003fostering} have been discussed to clarify the notion of meaningful work. All the aforementioned studies define meaningfulness as a subjective concept. Blustein et al. \cite{blustein2023understanding} categorized the core tenets of meaningful work into four segments:
(1) Organization-specific. It comprises the workplace ecosystem along with its practices and policies. It includes aspects like corporate social responsibility \cite{aguinis2019corporate}, and professional growth opportunities \cite{fletcher2019can}; 
(2) Social-context related. It deals with the interpersonal relationships existing in the working environment. Forging fulfilling relationships with coworkers \cite{fouche2017antecedents}, having positive workplace connections \cite{colbert2016flourishing, seppala2013social}, and a sense of belonging \cite{pratt2003fostering}, supportive colleagues \cite{bailey2016makes, martela2018autonomy, lysova2019fostering}, feeling appreciated by colleagues through direct or indirect gestures\cite{wrzesniewski2003interpersonal}, and being guided by competent leadership \cite{carton2018m} all aid in manifesting meaningful work; 
(3) Job/occupation design–related. It refers to the characteristics that accompany a role at the workplace. The job characteristics model by Hackman and Oldham cites attributes like autonomy, task identity, task significance, skill variety, and feedback as critical components associated with meaningful work \cite{hackman1976motivation}. Challenging work climate (in terms of workload, responsibility, and learning demands) \cite{kim2020thriving} and work-impact visibility \cite{grant2007impact, grant2007relational} are other relevant facets of job design; 
(4) Employment-related. It encompasses the various types of employment (like full-time, part-time, volunteer work, entrepreneurship, and so forth) and subsequent specificities in the form of job security \cite{arnoux2016perceived}, remuneration and workload distribution \cite{lips2020effect}.

Although there exist overlapping elements between decency and meaningfulness, like employment-related aspects of compensation or job security, they are two separate constructs. Decent work encompasses essential workplace conditions that every employee should have, while meaningful work is an ideal, embodying significance in employees' professional lives. Unlike decency, meaningfulness goes beyond mere material benefits and encompasses fulfillment and purpose in one's job.
Our distinction between job decency and job meaningfulness echoes Herzberg’s Two-Factor Theory \cite{hertzberg1959motivation}, where decency relates to hygiene factors (e.g., salary, working conditions) and meaningfulness to motivators (e.g., purpose, recognition). While we drew on more recent frameworks to reflect the evolving landscape of the future of work, this link highlights the theoretical continuity in understanding what drives job satisfaction.

\subsection{Job Satisfaction in Collaboration with AI}

Organizations are integrating AI technologies into their work environments to improve efficiency and reduce costs, all while ensuring consistent product quality \cite{russell2016artificial, miller2017robots, bhaumik2018ai}.
Yet, the complex challenges faced in today's business environment cannot be addressed solely by automated systems \cite{raftopoulos2023human}.
Effective solutions demand a synergistic collaboration between humans and AI, leveraging the unique strengths of both \cite{crandall2018cooperating, akata2020research, de2021ai, dellermann2021future}. 
Such a collaboration merges the advantages of human traits—such as perception, social skills, and creative problem-solving, with the benefits of AI, including speed, precision, adaptability, and flexibility \cite{braga2017emperor, you2018enhancing, esterwood2020human}.
While existing research on HAIC has looked into various aspects of working with AI-based systems \cite{lindvall2021rapid, han2024teams, qian2024take, xu2023comparing, kobiella2024if, xu2024makes, guo2024exploring, gu2024data}, they mostly focus on improving human-AI team performance. Less is known about the experiential outcomes of such a collaboration. Over the last few years, this research gap has been recognized by researchers and there has been a growing interest in exploring HAIC in the workplace beyond purely efficiency-focused results.

Walliser et al. \cite{walliser2019team} observed that adjusting role dynamics to consider humans and AI as teammates can foster greater emotional connection and teamwork. 
This was echoed by Sadeghian et al. \cite{sadeghian2024soul} that perceiving AI as teammates, invokes a greater sense of job meaningfulness, while AI as a superior decreases it. Furthermore, humans find their job more satisfying when they maintain significant autonomy and remain accountable for their actions.
Consistent AI responsiveness can support this autonomy by enabling humans to make choices and continue pursuing their goals, as highlighted by Mitchel et al. \cite{mitchell2021automated}.
According to Nyholm and Smids \cite{nyholm2020can}, robots can contribute to job meaningfulness by acting as "good colleagues" that learn and work constructively with humans, engage in pleasant and appropriate conversations, are respectful and supportive, and share work-related values. 
Smids et al. \cite{smids2020robots} further add that robots can enhance a sense of purpose at work by taking over repetitive tasks and leaving intellectually stimulating tasks for humans. They can positively influence growth opportunities by necessitating the acquisition of new skills. However, they might also decrease social interactions at work due to the presence of fewer human colleagues, leading to isolation and subsequently, reduced meaningfulness. While they can reduce appreciation and recognition by establishing the same equitable conditions for all, they can conversely enhance the social image associated with many occupational domains due to interacting with AI technologies. 
In line with such a twofold consequence, Waardenburg \cite{waardenburg2024human} emphasized the complex implications of automating straightforward customer service tasks within an organization. While enhancing the emotional support aspect of the role increased employees' sense of meaningfulness, the shift of simpler, "take-a-breather" tasks to AI negatively impacted their satisfaction due to the "psychological heaviness" associated with their remaining responsibilities.
As such, the perception of meaningful human control is an integral factor promoting job satisfaction.
Building on this idea of control, Hemmer et al. \cite{hemmer2023human} observed that AI delegation, regardless of human awareness, enhances self-efficacy and, consequently, task satisfaction. Their findings suggest that when individuals believe they are in control, their level of engagement and sense of satisfaction in their work increase.

The dynamics of support in the workplace also play a significant role.
In a separate study, Sadeghian et al. \cite{sadeghian2022artificial} observed that collaborating with humans tends to be more motivating and meaningful compared to working with an AI, regardless of the nature of the task. 
Likewise, Ossadnik et al. \cite{ossadnik2023man} investigated the beneficial effects of workplace assistance on job satisfaction through increased perceptions of organizational support. While they found no significant difference when help came from AI or conventional technology, the perceived organizational support was higher when help came from coworkers, compared to from AI.
Lastly, the integration of AI in workplaces also brings forth a fear of job loss that is not limited to the low-skilled workforce \cite{hirst2014does, agrawal2017expect, ivanov2017robonomics, davenport2018artificial}.
Bordot \cite{bordot2022artificial} studied the relationship between AI, robots, and job security using longitudinal data from 33 OECD countries\footnote{Australia, Austria, Belgium, Canada, Czech Republic, Denmark, Estonia, Finland, France, Germany, Greece, Hungary, Iceland, Ireland, Israel, Italy, Korea, Latvia, Lithuania, Mexico, Netherlands, New Zealand, Norway, Poland, Portugal, Slovak Republic, Slovenia, Spain, Sweden, Switzerland, Turkey, United Kingdom and the United States.}. He found that as automation technologies become more prevalent, the unemployment rate will tend to rise. However, this unemployment is not influenced by the education level of individuals.

While there exists some literature discussing the impact of AI on specific attributes of job decency (like job security \cite{nam2019technology}), and job meaningfulness (like skill variety \cite{findlay2017employer}), further research is required that considers multiple nuances of job satisfaction concurrently.
This gap in literature leads to our research question: "How will the introduction of AI in workplaces change the perception of job decency and meaningfulness across different occupational domains?"

\section{Method}
Our literature review showed that prior research on the impact of AI technologies in workplaces tends to focus more on objective outcomes than on employees’ subjective perceptions \cite{chui2015four, kolbjornsrud2016promise, mcclure2018you, bhargava2021employees}.
As such, we adopted an \textit{ethnographic futuring} approach \citep{Textor1995, EnglishLueck2021} to investigate how workers in different occupational domains imagine the future of job decency and meaningfulness under conditions of workplace AI. Ethnographic futuring, sometimes called anticipatory ethnography (see e.g., \cite{lindley2014anticipatory}), extends conventional ethnographic inquiry into speculative terrain by eliciting participants’ own projections, hopes, and fears about possible futures. In practice, we conducted semi-structured interviews \cite{lazar2017research} with IT, service, and healthcare workers, asking them to describe their current experiences of job decency and meaningfulness, and then to consider how these qualities might change in a future scenario where AI-driven systems shape or mediate their work. 
To analyze how participants situate their accounts temporally, we also draw on insights from \textit{temporal ethnography} \citep{Elliott2017, Dawson2014}, which highlights how orientations toward past, present, and future are socially and culturally constructed.
We explored participants’ perceptions of job decency and meaningfulness in both the \textit{present} and the \textit{future} of work with AI, as we believed that both current work experiences and anticipated changes could jointly shape their attitudes, expectations, and concerns.

This approach, compared to other forms of anticipatory ethnography that use design fiction artifacts as ethnographic probes \cite{lindley2014anticipatory}, has limitations such as unconstrained and uneven imaginaries shaped by participants’ existing familiarity with AI. Nevertheless, the combination of ethnographic futuring and temporal ethnography enables a grounded exploration of how people imagine the future of work with AI, while situating those imaginaries in relation to present lived experience and temporal orientation.

\subsection{Participants}
Twenty-four participants (7 female, 17 male, 0 diverse) from twenty-two different organizations across three professional domains, IT (n=8), service-based (n=8), and healthcare (n=8), aged between 24\textendash{}46 (M=31.38, SD=4.46), recruited through social media and word-of-mouth, participated in this study between January-April 2024.
The aim was to include participants from different roles and organizations within each domain, which required reaching individuals across multiple professional networks.
Furthermore, we selected an equal number of participants per domain to ensure balanced representation across IT, service, and healthcare.
We selected these three domains because they represent a broad spectrum of work - technical (IT), relational (service-based), and care-oriented (healthcare), all of which are all being reshaped by AI \cite{sigov2024emerging, limna2023artificial, shaheen2021applications}.
The technical domain primarily involves task-oriented work, while the relational domain thrives on interpersonal interaction and social connection. The care domain, in turn, is deeply human-centered, focusing on empathy and emotional support. 
As such, we believed that comparing them would allow us to examine how AI affects perceptions of job decency and meaningfulness across contrasting forms of work.

Two participants from the service domain and two from healthcare belonged to the same organization, but all held different roles.
The first group, IT domain, comprised software engineer(3), machine language engineer(2), data analyst(1), deep learning engineer(1), and front-end developer(1).
The second group, service-based domain, comprised waitstaff(4), floor-leader(2), bartender(1), and cook(1).
The third group, healthcare domain, comprised intensivist(2), general practitioner(1), internist(1), microbiologist(1), otorhinolaryngologist(1), pediatrician(1), and psychiatrist(1).

All IT participants reported having prior exposure to AI, specifically through large language models like ChatGPT\footnote{https://chat.openai.com/} or GitHub Copilot\footnote{https://github.com/features/copilot}, using them in their everyday work. Two service-domain participants mentioned having \textit{"tried out ChatGPT"} out of curiosity, but used it infrequently thereafter. Three healthcare participants were acquainted with AI technologies used in Radiology, although they had not personally applied them in their practices. Another healthcare participant cited three AI-based tools (Elicit\footnote{https://elicit.com/welcome}, Julius AI\footnote{https://julius.ai/}, and Paper Digest\footnote{https://www.paperdigest.org/}) that he regularly used to further his medical research pursuits.
Participants were based in Europe(n=14), Asia(n=9), and North America(n=1).
Within each domain, IT participants came from Europe(5), Asia(2), and North America(1), service-based participants from Europe(7) and Asia(1), and healthcare participants from Asia(6) and Europe(2).
Each participant was interviewed separately via Webex\footnote{https://www.webex.com/}. Before the interview, they were asked to sign a consent form outlining anonymization of participants' data and the use of recordings solely for research purposes. The interviews lasted one hour each, and participants received €12 in compensation.
The study was approved by the ethics committee of the University of Siegen before commencement. 

\subsection{Procedure}
Interviews consisted of two segments: one focusing on their current experiences (named, present scenarios), and another envisioning future experiences (named, future scenarios). Each segment had three parts: introductory, decent work, and meaningful work.

In the \textit{present} introductory part, we asked the participants to describe a typical working day and tell about their work environment, technologies used, and satisfying and dissatisfying job aspects. For the \textit{future} introductory part, we asked them to imagine an AI of \textit{their choice} and describe what working with it might look like in the future.
This open framing was intentional, allowing participants to draw on their own experiences, expectations, and imaginaries of AI rather than being constrained to a predefined system or capability.

\begin{figure}[h!]
    \centering
    \includegraphics[width=1.00\linewidth]{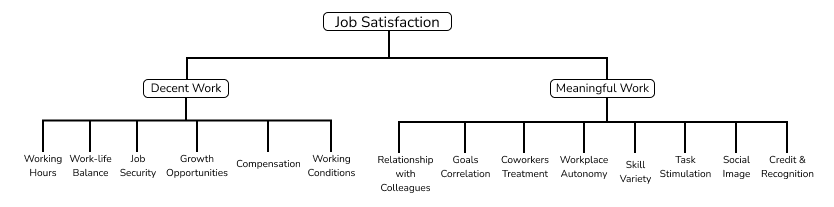}
    \caption{Determinants of Job Satisfaction (Decency and Meaningfulness) and their underlying attributes.}
    \label{fig:decency_meaning} 
    \Description{Image consists of a header box named "Job Satisfaction", below which are two boxes showing its two determinants - "Decent Work" (which has six underlying attributes) and "Meaningful Work" (which has eight underlying attributes).}
\end{figure}

For the decency part, in line with \cite{pereira2019empirical, office2014global}, we asked them questions related to their working hours, work-life balance, job security, growth opportunities that their workplace provides, compensation, and working conditions (Figure \ref{fig:decency_meaning}). In the \textit{present} segment, participants answered these questions \textit{based on their current experiences} (present scenarios), while in the \textit{future} segment, they responded \textit{based on their speculation of the AI} at their workplace (future scenarios).

Likewise, for meaningfulness part, in line with \cite{aguinis2019corporate, fletcher2019can, colbert2016flourishing, wrzesniewski2003interpersonal, hackman1976motivation, kim2020thriving, arnoux2016perceived, lips2020effect}, we first asked participants a general question on their job fulfillment, followed by questions on inter-colleague relationships, alignment of goals with those of their organizations, coworkers’ treatment, task autonomy, skill variety  (encompassing diverse tasks with varying levels of challenge), task stimulation (perceiving tasks as intellectually engaging or monotonous), social image, and workplace recognition (Figure \ref{fig:decency_meaning}).

\section{Results}
The interview recordings were transcribed, verified, and anonymized by the first author. 
For the analysis, we followed the emergent coding method followed by thematic analysis \cite{lazar2017research}. We used MAXQDA\footnote{https://www.maxqda.com/} software to code our data. 

Participants from IT domain envisioned their AIs as assistants providing them with a \textit{"helping hand"} that would highlight action points (n=4), plan upcoming schedules (n=3), attend and summarize meetings (n=3), brainstorm solutions (n=3), understand development context to prompt in-code suggestions (n=3), summarize unread emails (n=3), prepare discussion contents before meetings (n=2), develop software (n=1), engage in informal conversations (n=1), and be a constant work companion (n=1).
AIs were described as physical entities (robots, n= 3), digital components (plugins, n= 6), or avatars in the virtual space (Metaverse avatars, n= 2), with various interaction modalities, including verbal, textual, visual, neural, and haptic.

Participants from service domain envisioned their AIs as tools that could ease their everyday burden by sharing the workload. 
AIs would present menu and take orders (n=4), speak with customers (n=3), serve food (n=3), generate end-of-day activity reports (n=3), manage inventory (n=3), highlight purchase requirements (n=2), do cleaning (n=2), capture attendance and work time (n=1), and estimate daily customer volume in advance (n=1).
AIs were described as physical entities (androids/robots, n=5), or digital components (software programs, n=4), with various interaction modalities, including verbal, textual, visual, and touch.

Participants from healthcare domain envisioned their AIs in the form of assistants. 
AIs would fetch patient history (n=5), patient monitoring (n=3), preemptive identification of health issues (n=3), identifying abnormalities in reports and scans (n=3), scan comparisons (n=3), conducting scans (n=2), documentation (n=2), highlighting overlooked details or concerns while specifying medication (n=2), diagnosing and creating treatment plans (n=2), note-taking during consultations (n=1), and post-discharge patient follow-ups (n=1).
AIs were described as physical entities (robots, n=3), or digital components (software programs, n=8), with various interaction modalities, including verbal, textual, visual, and touch.

As such, across domains, AIs were envisioned to handle both knowledge-based tasks and physical labor. IT and healthcare participants anticipated AI assisting with knowledge-intensive, error-sensitive tasks, such as coding or diagnosing health conditions. In contrast, service participants envisioned AI handling a mix of physical tasks, like cleaning or serving, alongside administrative duties, such as managing inventory.

\subsection{Job Decency Results (Present versus Future)}
Qualitative insights revealed diverse perceptions among participants regarding all aspects of job decency, emphasizing both threats and opportunities in collaboration with AI, which are listed below.

\subsubsection{Working Hours}\hfill\\
\textbf{IT workers:}
In present scenarios, participants reported having 39\textendash{}40 weekly hours, although the \textit{actual time worked} was slightly less. 
Six found it adequate, citing flexible timing and dedicated idle time, while two found it prolonged and excessive. 
In future scenarios, five felt their working hours would be reduced with AI sharing the workload. One believed that hours would change only if they are \textit{"supposed to do the same amount of work [as before] after introducing the AI"} (IT-P8), while another deemed that they would still be expected to spend the same hours, as before, on newly added tasks. One highlighted a dilemma regarding producing higher output by working the same hours as before, or reducing hours to only match the previous output volume.
 
\textbf{Service workers:}
In present scenarios, participants mentioned working 20\textendash{}60 hours per week. Four found it standard, and two found it rewarding for their earnings. However, two considered it to be cumbersome due to the physical nature of work: \textit{"We have a lot of physical stress, it's not good. It will cause [health] problems later."} (SV-P8).
In future scenarios, six believed there would be no change in their working hours since the arrival of customers is not dependent on AI.
Two said they would be dissatisfied with their new hours - while one believed their reduced hours would lower their wages, another felt the hours would increase, since mishaps by AI would be more severe, and on a larger scale than those made by humans.

\textbf{Healthcare workers:}
In present scenarios, participants mentioned working 50\textendash{}75 hours weekly. While none considered the hours appropriate, they felt there was no alternative due to \textit{"gross deficit of healthcare professionals"} and vast doctor-patient ratio: \textit{"It cannot be compared to another sector. There is a service which needs to be provided to patients, so there's no other way."} (HC-P6).
In future scenarios, five felt working hours would be reduced with AI sharing the workload. While three believed hours would remain unchanged since \textit{"they would still be responsible for patients"}, they felt it would be less strenuous than before due to AI assistance.

\subsubsection{Work-life Balance}\hfill\\
\textbf{IT workers:}
In present scenarios, all eight participants found their work-life balance satisfying due to proper workload management, \textit{"no obligation to work overtime"} (IT-P4), hybrid work-mode, and flexible working hours: \textit{"When I have flexibility in my work hours, then I have flexibility in my personal life."} (IT-P8). 
In future scenarios, four believed that the time cutback would improve their work-life balance, while three could not foresee any differences due to the need to be present to drive and complete the work, or changing expectations at the workplace: \textit{"... when everyone's productivity is boosted, it will change the expectations - the nature of work would be different!} (IT-P5).

\textbf{Service workers:}
In present scenarios, seven participants were satisfied, citing a clear separation between their working and non-working hours: \textit{"my work-stress ends with my shift"} (SV-P5). One participant desired better workload management to avoid being \textit{"too tired after returning from work"}.
In future scenarios, two felt their work-life balance would be healthier with AI sharing their workload, while six expected no change since the nature of their industry is such: \textit{"The balance doesn't depend upon what's happening within the shifts, but it is the nature of this type of job."} (SV-P4).

\textbf{Healthcare workers:}
In present scenarios, none believed they had a healthy work-life balance due to the nature of their work: \textit{"We are [available] 24x7 on call. And I'm not hesitant to answer, because one call can save one life."} (HC-P7). Participants reiterated that this aspect cannot be compared with other professions where the \textit{"concept of holidays"} exists. They believed balance in work-life is synonymous with not working overtime (n=6), having two days off per week (n=5), and having sufficient leisure time (n=4).
In future scenarios, five believed that reduced working hours would improve their work-life balance. Three felt that \textit{"balance would remain the same"}, but \textit{"quality of work-life would improve"}: \textit{"With AI, I can [access patient's current health status] quickly and proactively, and I will get some peace of mind at home."} (HC-P7).

\subsubsection{Job Security}\hfill\\
\textbf{IT workers:}
In present scenarios, two participants did not feel secure due to the economic crisis. Six felt secure due to their vital roles, favorable employment laws, organizational support, or meeting expectations. 
In future scenarios, five felt sustained learning and development are paramount for employment stability. Fear of layoff was non-existent amongst four who \textit{"expect[ed] to have the brain"} in human-AI teaming, or were adamant that \textit{"AI is not here to take our jobs"}. However, three felt job security would decline as AIs become more competent with time. 

\textbf{Service workers:}
In present scenarios, six felt secure due to their work proficiency, the lack of \textit{"[competent] human resources"} (SV-P4), and organizational support. One mentioned feeling insecure as they were newly hired.
In future scenarios, four participants anticipated job security, citing pivotal roles, need for human resources, skepticism about AI advancement in their lifetime, and prevalent organizational support. One, being newly hired, anticipated being the first to lose their job, and two more predicted being replaced by AI since the \textit{"job is not really [mentally demanding], but just manual work"} (SV-P3).

\textbf{Healthcare workers:}
In present scenarios, all eight participants felt secure due to the vast doctor-patient ratio \textit{"there's already a severe dearth of doctors in the profession"} (HC-P1), and the consequent shortage of specialists, favorable employment laws, and organizational support. 
In future scenarios, four felt their jobs would be secure since \textit{"the brains [in human-AI teaming] will be doctors"} (HC-P2), and patient responsibility would still rest with them. Although another expected their position to be secured due to their government contract, they felt there would be job loss on a wider scale. Meanwhile, three felt job security would decline. One believed their general practice background would make their position vulnerable, while another felt insecure due to the eventual reduction in demand for doctors. Another anticipated job saturation in city-based hospitals that could afford AI, and considered moving to rural areas for employment: \textit{"If AI [comes], then a doctor can only go to the village and do private practice."} (HC-P7).

\subsubsection{Growth Opportunities}\hfill\\
\textbf{IT workers:}
In present scenarios, all mentioned having abundant learning avenues, but two said the topics hinge on market trends. With various growth prospects, participants felt responsible for utilizing them: \textit{"It's on me to motivate myself to take this budget or to take this time and do something meaningful instead of just wasting it!"} (IT-P4).
In future scenarios, all eight believed that the need to learn new skills would be perpetual due to evolving demands, and there would be no dearth of opportunities to grow. Two participants stated that the scope of growth would expand since AI would do repetitive tasks and they would get \textit{"more challenging [problems] which are not solved yet"}. Furthermore, effective AI utilization will shape one's development prospects, since everyone has access to the technology: \textit{"In the end, everyone has this AI. So they all have the opportunity to be perfect but that won't happen because ultimately how you take suggestions or help from AI and apply it is up to you. So that will differentiate you [from] your colleagues."} (IT-P8).

\textbf{Service workers:}
In present scenarios, six felt interacting with customers helps improve their interpersonal skills. While two mentioned that evolution within the service sector is uncommon: \textit{"it's an occupation where you do something repeatedly and you cannot do something else"} (SV-P8), another highlighted organizational training measures for skill development and promotions.
In future scenarios, five felt growth opportunities are not dependent on AI since their interaction with customers would remain the same, while three believed opportunities would be restricted because AI might replace humans in the service sector.

\textbf{Healthcare workers:}
In present scenarios, six were satisfied with the numerous training sessions, hands-on workshops, fellowships, and research opportunities to upskill. Two were not content due to limited opportunities for research or scaling fast.
In future scenarios, four envisaged growth opportunities to increase since they would \textit{"get to learn about new technologies and use them in different ways"}. One further believed that information would be readily discoverable and accessible to all: \textit{"I feel education is highly commoditized and the publishing houses are strangling the flow of information. With AI coming in, information would be really quick, and affordable, and I think that can be utilized by people to upgrade themselves."} (HC-P6).
While one believed the scope of growth outside the domain of specialists would cease to exist, three felt growth opportunities are independent of the presence of AI: \textit{"I think an AI would not in any way affect what a doctor needs to learn to provide proper patient care."} (HC-P1).

\subsubsection{Compensation}\hfill\\
\textbf{IT workers:}
Remuneration, in present scenarios, was benchmarked against geography-specific market salaries for similar jobs, with five participants being satisfied, and three dissatisfied. 
In future scenarios, three articulated that abundant growth opportunities would enhance their competencies and increase their professional value. However, one could not envision any changes since using AIs will be commonplace: \textit{"It will become normal to use AI tools; so I don't think that will change the compensation the companies will provide."} (IT-P1). Two participants deliberated whether it made sense to reduce their salary if they had to work fewer hours than before, and another believed that organizations would offset AI maintenance costs by decreasing the payroll: \textit{"From a company's perspective, if it is spending on something which is doing [part] of the job, and also employing a human who is doing the remaining work, [the budgetary] impact is going to be on the on the human part, not on the AI part"} (IT-P6).

\textbf{Service workers:}
In present scenarios, six were content with their pay as it covered daily expenses, although all desired higher earnings: \textit{"as humans, you always aspire for more"} (SV-P1). Two felt underpaid given their efforts and the market salary for other jobs.
In future scenarios, five believed their compensation would not be reduced since it is safeguarded by labor laws, and \textit{"depends upon tips from customers"}. One anticipated improved pay for providing better value to the restaurant: \textit{"I'd get more salary because the organization will have to spend a lot to [have multiple] robots. However, I can charge them much less for availing my services."} (SV-P7). Two felt their salary could be reduced based on the volume of activity in the restaurant.

\textbf{Healthcare workers:}
In present scenarios, seven felt they were not well compensated, considering the extensive years of education invested to attain their positions, and the extended time commitment demanded by their roles. One further remarked that discussions about salary increments for doctors are met with disapproval, as lawmakers argue that medical profession's essence lies in altruism, not profit.
In future scenarios, two participants felt AI would not influence their remuneration since \textit{"they would still be the primary caregiver of the patient"}. Three believed that hospitals would offset the cost of acquiring AI by reducing their salaries instead, and two felt that the reduced number of incoming patients would affect their income: \textit{"When the patient is getting instant relief through AI-doctors on their mobile phones, there will be a trend for a good number of patients to use this system and to try to get their diagnosis."} (HC-P4).

\subsubsection{Working Conditions}\hfill\\
\textbf{IT workers:}
In present scenarios, all spoke positively about their working conditions, citing aspects like open-office space, flat hierarchies, post-work meetups, and supplementary amenities. 
In future scenarios, all felt the work environment and culture for humans would not change, but two of them expected the work facilities with AI to become highly sophisticated, and costly to maintain. While one believed that multiple companies will share single AI facilities through \textit{"smaller localized hubs"}, another hypothesized a neural interaction-based mode of working: \textit{"[Previously,] if I want[ed] to design a system, [and fell asleep thinking], then next morning I [would have had] to [recollect and] put it [into action]. But now with AI assisting me, I can probably just think of it and it's done!"} (IT-P5).

\textbf{Service workers:}
In present scenarios, all eight participants felt contented with their working conditions. They mentioned having access to clean uniforms (n=8), complimentary beverages and surplus food (n=8), organization-sponsored outings (n=3), and \textit{"a big window with fresh air"} (SV-P3). 
In future scenarios, two participants felt they would have \textit{"less stress with orders"} since the \textit{"burden will be reduced"} with AI assistance. While four anticipated no change in the working culture for humans, two believed the human element would be lost: \textit{"AI will manage everything, from food preparation to giving services to customers."} (SV-P6).

\textbf{Healthcare workers:}
In present scenarios, four participants were satisfied with their working conditions, as they had access to all necessary medical tools. Three expressed dissatisfaction citing \textit{"inadequate [hospital utilities] to provide quality healthcare services"} (HC-P4), and the looming sense of bereavement: \textit{"Patients are sick and dying, there are grieving parents, so it's not a happy environment, and it gets to you after a point."} (HC-P8). All participants enjoy post-work meetups with colleagues, even though they are infrequent, with everyone being overworked.
In future scenarios, all felt the working environment would be \textit{"a lot less stressful"} with AI absorbing some of their workload. 
While six felt the work culture for humans would not change, two believed the way humans interact with each other would transform: \textit{"Fewer people are going outside to play nowadays. So maybe [human-human interactions] would happen virtually, and while that might seem alien now, it would be normal for that period."} (HC-P6).

\subsection{Job Meaningfulness Results (Present versus Future)}
Qualitative insights from interviews revealed diverse perceptions among participants regarding all aspects of job meaningfulness. We started this part of the interview with a general question on meaningfulness, followed by further questions on specific elements of meaningful work.

\subsubsection{Job Meaningfulness}\hfill\\
\textbf{IT workers:}
In present scenarios, all found their jobs meaningful. Having significant impact (n=5) \textit{"I feel like what we are doing is different, definitely affecting our users and making their lives better"} (IT-P1) and growing professionally (n=3) \textit{"meaningful means getting to learn something new"} (IT-P6) were named as sources for meaning.
In future scenarios, six participants believed their work would still be worthwhile since humans would use the AI as \textit{"a tool"} to be more efficient, while two felt otherwise since their jobs could be taken over by AI.

\textbf{Service workers:}
In present scenarios, five found their jobs meaningful citing the opportunity to meet and converse with different people (n=4), and devotion to their craft (n=2). Three did not find their jobs meaningful citing it to be routine.
In future scenarios, five felt their jobs would remain meaningful since they would still get to meet and interact with different people and perform their obligations with dedication, while three believed their jobs would still be boring.

\textbf{Healthcare workers:}
In present scenarios, all found their jobs meaningful because they \textit{"save human lives"} and create \textit{"impact on the society"}. However, for one participant, hobbies brought him more fulfillment than his work.
In future scenarios, seven participants believed their work would remain meaningful since they would still be treating human beings and would still \textit{"have the responsibility of another life"} with them. One, however, feared his role could become obsolete: \textit{"As long as I have the main role, I am still going to talk to patients and treat them. But if AI does diagnosis and prescribes medication, then there is no place for me."} (HC-P5).

\subsubsection{Relationship with Colleagues}\hfill\\
\textbf{IT workers:}
In present scenarios, inter-colleague interactions were appreciated by seven participants, with in-person engagement preferred to remote. Participants mentioned aspects like open dialogue, mentoring, empathizing with work issues, and socializing.
In future scenarios, the collective accord was that interpersonal relationships would be more profound as conversations shift away from work-related inquiries to personal and informal topics: \textit{"Maybe you will not interact about your work; maybe you can interact about other things! You can ask how was your day and all that."} (IT-P1). Four said that due to humans' inherent social nature or since workplace ownership still rests with humans, they would prefer interacting with humans to AIs.

\textbf{Service workers:}
In present scenarios, seven were contented with their inter-colleague interactions alluding to the opportunities to learn from each other (n=5) and engage in casual conversations (n=4).
In future scenarios, three participants imagined the volume of interactions to increase with AI sharing the workload: \textit{"[Interactions] will get better and frequent because we will have more time to speak with each other."} (SV-P8), while five felt inter-colleague exchanges would decrease since there would be fewer human coworkers. One further expressed reluctance about such a future: \textit{"The point of working [together] is because you can talk with them and laugh together. But you cannot do these things with a machine. And I think that is so horrible!"} (SV-P3).

\textbf{Healthcare workers:}
In present scenarios, all eight participants highlighted that interacting with colleagues is a critical aspect of their profession: \textit{"It's an absolute teamwork. Nobody can do the role of a medical professional on a solo basis."} (HC-P1). Learning from others (n=8), getting a second opinion on differential diagnosis (n=6), and having lighthearted interactions (n=3) made inter-colleague interactions engaging. 
In future scenarios, four believed there would be no change in their inter-colleague interactions since they would still collaborate with their coworkers to take better care of their patients. One further added that humans would prefer interacting with humans to AIs: \textit{"Even if a robot is right and much more accurate than me, people will never like to listen about [the diagnosis] from them."} (HC-P8). With AI sharing the workload, two believed that \textit{"communication [with humans] would be elevated to something more substantial"} since they would get more time to bond with their colleagues. Contrarily, two others imagined the volume of conversations to go down: \textit{"Except for the information held by specialists, all other information would be available [with AI]. So communication would be reduced."} (HC-P6).

\subsubsection{Goals Correlation}\hfill\\
\textbf{IT workers:}
In present scenarios, five felt their goals are aligned with those of their organization since their aspirations are accounted for by higher management while planning the roadmaps. In contrast, one reflected on the organizational tendency to coordinate employee goals with market trends and project availability.
In future scenarios, four felt their goals would align with the organization since it would still involve upskilling, being more efficient, and creating impact. 
One lamented the obligation to align with only AI-centered goals: \textit{"It's not [our] own goals because AI is forcing us to move in one particular direction. If someone isn't interested in AI, he cannot exist in the industry [any] longer."} (IT-P3). Two stated this aspect is unrelated to AI.

\textbf{Service workers:}
In present scenarios, five considered their goals to be aligned with those of their employer restaurants since they work in agreement due to their passion for food and serving the customers. One participant, however, could not identify with the restaurants' goals stating: \textit{"the food is not that good"} (SV-P4).
In future scenarios, none could imagine AI influencing this aspect of their work. Participants aligned with organizational goals felt they would share the common passion nevertheless, while those disconnected from the goals believed they would continue to remain so.

\textbf{Healthcare workers:}
All participants mentioned their institutional goal as optimizing patient care. In this regard, in present scenarios, seven felt their goals are aligned, although three mentioned that situational circumstances, like limited resources, poor infrastructure, and strict hierarchy, prevent them from realizing these goals. One did not identify with their organization's goal, since they are more inclined towards research.
In future scenarios, three participants felt they would be better equipped to achieve the goal of optimizing patient care since \textit{"resources will be better organized with AI"}, while four stated this aspect is unrelated to AI.

\subsubsection{Coworkers Treatment}\hfill\\
\textbf{IT workers:}
In present scenarios, all reserved high praise for the treatment meted out to them by their colleagues. They found them to respect boundaries and be fair, transparent, nurturing, and treat them as equals.
In future scenarios, seven mused that human behavior will be independent of technological advancements. Four believed that workplace relationships would remain unchanged since AI would be used by all and exist only to assist them. However, one mentioned that his colleagues might perceive the AI as the principal executor of tasks and alter their behavior towards him: \textit{"If someone perceives like [I am] not doing anything [and] it is all done by the AI, [then he will think] why should [he] talk to [me]? [He] can simply go to the AI!"} (IT-P6).

\textbf{Service workers:}
In present scenarios, all eight participants found coworkers friendly, respectful, and non-discriminatory.
In future scenarios, seven asserted that morality is not reliant on AI, while another imagined exemplary treatment from the impartial AI: \textit{"It will be even better because now the colleague is the robot, so it'll be super fair"} (SV-P4).

\textbf{Healthcare workers:}
In present scenarios, six mentioned that their coworkers were friendly, cooperative, fair, and transparent. Two highlighted the workplace tendency to show deference to seniority instead: \textit{"It's all about authority. You fear the hierarchy. It's not about the person."} (HC-P8).
In future scenarios, four considered human behavior to be an intrinsic element that remains unaffected by AI, while three believed there would be no change in the treatment meted out to them since \textit{"the nature of the work doesn't change"} and AI is \textit{"only assisting"} them. One participant felt the respect they would earn from their colleagues would lessen since AI would make everyone equally adept in their respective roles.

\subsubsection{Workplace Autonomy}\hfill\\
\textbf{IT workers:}
In present scenarios, seven participants were content with the autonomy they exercised, even if they had to perform some predetermined tasks. However, one voiced a sense of helplessness: \textit{"You feel that your knowledge is not being used to the fullest, because if you know that there is a better solution, but you are told that - no, you have to stick to this, then you [do not feel] motivated."} (IT-P6).
In future scenarios, seven anticipated retaining their autonomy, if not augmenting it, as they would maintain control and tailor the AI to meet their needs: \textit{"AI will probably never be critical decision makers because humans will still be the decision maker ... the law cannot get hold of AI, someone has to be accountable."} (IT-P5).

\textbf{Service workers:}
In present scenarios, all were satisfied with their autonomy, despite having tasks assigned by superiors.
In future scenarios, seven expected to retain their autonomy, since AI would only be reducing their workload, while another believed the autonomy would increase by delegating unwanted tasks to AI without concern: \textit{"With human colleagues, you cannot always tell them - do this, do that. But you can tell the robot to do anything that you don't want to. And there's no problem with that!"} (SV-P4).

\textbf{Healthcare workers:}
In present scenarios, all participants attributed the ability to exercise autonomy in the medical profession to years of experience and seniority. Seven were satisfied while one longed for a \textit{"bit more autonomy"} than what they have.
In future scenarios, half of the participants imagined retaining their autonomy since they would still be the primary caregivers of the patient, while the other half anticipated losing some of their autonomy with the AI making decisions on their behalf. Even then, one felt this loss of autonomy would enable them to devote their time to something else.

\subsubsection{Skill Variety}\hfill\\
\textbf{IT workers:}
In present scenarios, all eight participants remarked that working on multiple technologies and solving novel problems were challenging. Performing background research to handle complex systems, translating vague requirements into actionable points, shouldering decision-making responsibilities, and minimizing downtime for critical systems are further instances of participants navigating through a myriad of challenges. Nonetheless, such challenges are what motivates them in their daily work: \textit{"It's something which excites me!"} (IT-P6).
In future scenarios, all believed that the evolution of work would elevate organizational expectations, with new problems arising: \textit{"When the tool advances, companies will come up with more challenging [tasks]."} (IT-P4). According to them, \textit{"challenges would always be there"}, but now they would have a \textit{"helping hand"} at their disposal.

\textbf{Service workers:}
In present scenarios, four considered their tasks to be challenging, while the remaining four had grown accustomed to the demands of their work. Challenges included communication across languages, physical demands, and managing customer rush.
In future scenarios, five participants expected their tasks to be more manageable with AI sharing the workload, while three felt co-existing with AI would bring its own set of challenges. 

\textbf{Healthcare workers:}
In present scenarios, all eight participants said that the demands and expectations from their roles made their everyday routine challenging. Making life-and-death decisions (n=5), complexity of human anatomy (n=4), administrative tasks (n=3), and research work (n=1) motivate them to be at their best.
In future scenarios, four participants felt their tasks would be challenging since critical decisions would still be under their purview, with AI only assisting them to be more confident with their diagnosis. While one believed that \textit{"adapting to the AI will be a challenge on its own"}, another asserted that \textit{"there would be a new set of problems that would arise with the advances in technology"}. Two felt the level of challenges would be lowered due to AI's help.

\subsubsection{Tasks Stimulation}\hfill\\
\textbf{IT workers:}
In present scenarios, all eight participants considered the challenges they encountered to be intellectually stimulating. However, some days would have their fair share of monotony: \textit{"It's not always stimulating, sometimes you have to work on boring stuff, but it's sufficient."} (IT-P4).
In future scenarios, six participants emphasized that the novelty of unfamiliar problems would make solving them intellectually stimulating since AI would tackle all monotonous tasks. However, two stated that their mental engagement would be reduced with AIs sharing their workload.

\textbf{Service workers:}
In present scenarios, four participants labeled their tasks monotonous, with little to learn after initial instances, while three felt \textit{"the interactions with customers make it stimulating"}.
In future scenarios, three participants thought integrating AI into their work would make it more engaging, while four felt the cognitive engagement would reduce and routine tasks would become \textit{"even more boring"}. For one participant, intellectual stimulation remains constant, since AI would \textit{"reduce only the physical aspect of work"}.

\textbf{Healthcare workers:}
In present scenarios, seven mentioned that every incoming patient represents \textit{"a unique case"} which makes diagnosing them intellectually stimulating. However, one participant confessed that even though \textit{"the work itself is stimulating"}, they are no longer overwhelmed after many years in their role.
In future scenarios, four expected their cognitive engagement to lessen with AIs helping them in their daily tasks. One mentioned that working with AI would be intellectually stimulating, and another anticipated novel and engrossing tasks since all \textit{"manual and repetitive work"} would be taken over by AI. Two believed interpreting the information from AI would require critical thinking on their part.

\subsubsection{Social Image}\hfill\\
\textbf{IT workers:}
In present scenarios, seven participants reported a positive image portrayal due to their organizations' reputation, acquired soft skills, and \textit{"solving difficult tasks"}.
In future scenarios, five envisaged no change in public perception since AIs would be ubiquitous and only exist to assist humans: \textit{"If we go back 40 years, laptops were not common. So people would think - okay, he's using a laptop, so he's getting all the help! But now, since we know that everyone is using [laptops], no one thinks like that. It will be same with AI."} (IT-P6). 
While one believed their social status would decline due to the \textit{"misconception of the public"} that AI does their work, another was concerned about their work image being affected by AI-driven social interactions: \textit{"If [my] AI posts something offensive, that will definitely impact my image."} (IT-P5). 

\textbf{Service workers:}
In present scenarios, five believed they portray a positive social image since their job makes them financially independent (n=5), overcome boundaries (n=2): \textit{"For many years I worked as a teacher, and many people thought I cannot work [anywhere else]. But now I am working in a restaurant. And they are very surprised and happy for me."}, and recognized professionally (n=1). Two participants considered their social standing to be low, with their work being \textit{"considered a physical job that anyone can do"}, and irrelevant to their educational background.
In future scenarios, all eight participants believed their social standing would rise, since \textit{"everyone would know that [they] can work with AI"}.

\textbf{Healthcare workers:}
In present scenarios, all felt everyone acknowledged their indispensable role in society, with three highlighting the \textit{"excessive reverence"} associated with the medical field.
In future scenarios, six participants envisaged no change in public perception since their nature of work would remain the same, and they would carry on being an integral part of society. While one believed \textit{"the aura about doctors would fade with AI suggesting medications"}, another felt that the respect shown to doctors would be \textit{"much more genuine once the whole illusion of what a doctor does goes away"}.

\subsubsection{Credit and Recognition}\hfill\\
\textbf{IT workers:}
In present scenarios, all eight participants revealed they receive due appreciation for their efforts and feel their work is acknowledged, although, not everything warrants commendation: \textit{"There isn't acclamation for everything you do, it doesn't happen in professional life."} (IT-P8). 
In future scenarios, six believed they would be appreciated due to their enhanced competencies, normalization of AI-usage \textit{"we won't look down on someone using AI because we are all using it"} (IT-P6), workplace culture, or lack of liability within AI systems \textit{"AIs won't be credited, the accountability and ownership still stays with us"} (IT-P5). However, two felt they would be appreciated less due to the tacit perception of being guided by AI.

\textbf{Service workers:}
In present scenarios, all participants mentioned their contributions being recognized and appreciated by superiors (n=8), and colleagues (n=4).
For the future scenarios, four anticipated reduced recognition since they would be competing with AI, while four expected no change since their work obligations would remain the same, with AI only reducing stress.

\textbf{Healthcare workers:}
In present scenarios, six participants mentioned they receive appreciation from patients and their families (n=4), as well as from their colleagues (n=4) when recipients recover. However, two participants, from critical care units, considered their efforts to be unrecognized by patients: \textit{"Critical care is a thankless job. Patients come in very sick; we treat them and then they are shifted to regular wards. So you don't get to interact with them when they get better."} (HC-P2).
In future scenarios, half of the participants believed they would still be appreciated since \textit{"the human aspect is irreplaceable"}, while the rest expected to be credited less due to the diagnostic assistance from AI: \textit{"A whole lot of respect comes from remembering minute details of patients that [other doctors] might have missed. So, if this ability to recollect is delegated to AI, obviously the patient will benefit since all doctors will get the same notification from AI. But then, the respect will be nothing great because we all remember the same thing. So that one specific showmanship of talent of a particular doctor would reduce."} (HC-P3).

\subsection{Summary of Results}

\begin{figure}[h!]
    \centering
    \includegraphics[width=1.00\linewidth]{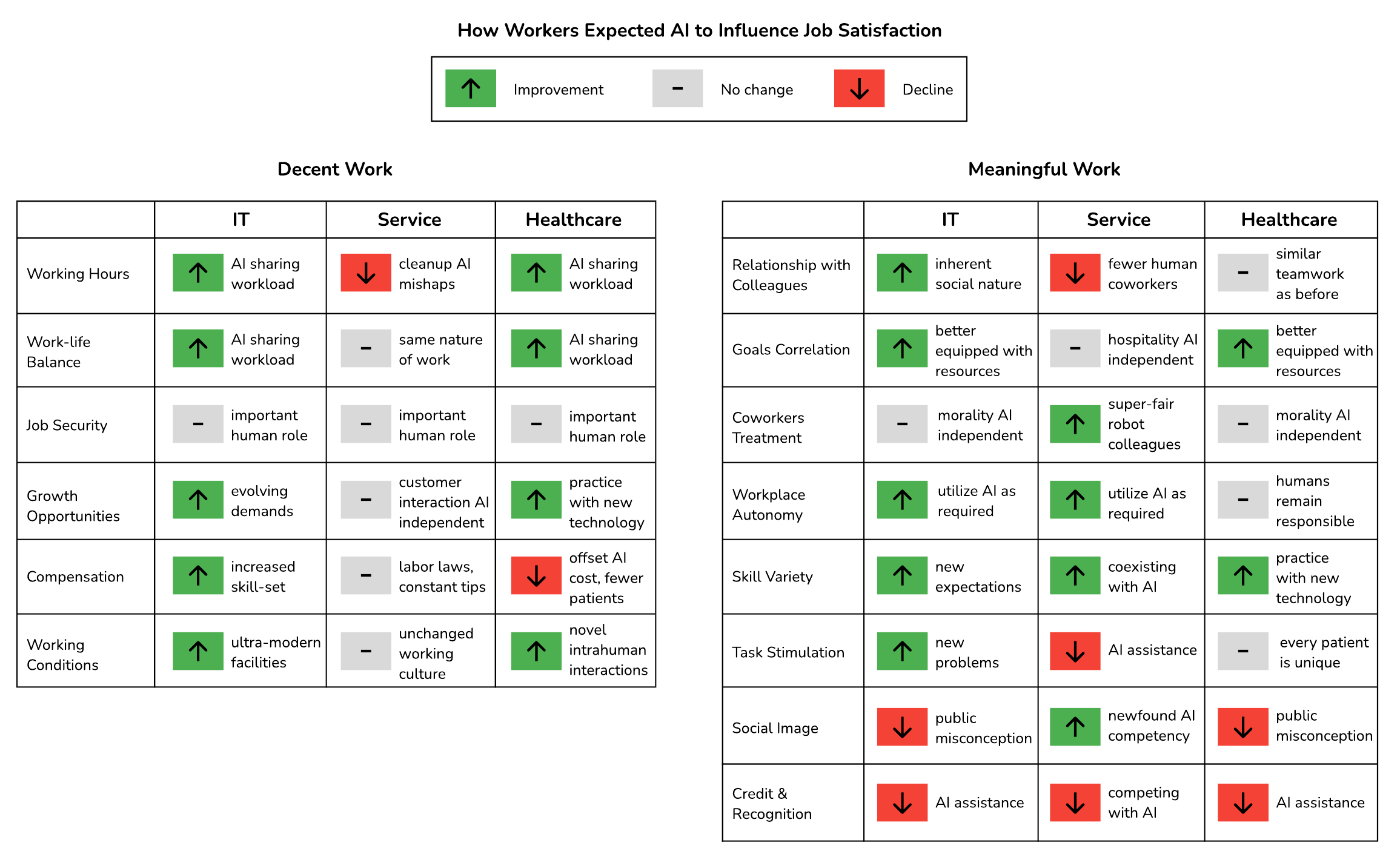}
    \caption{Anticipated impact of AI on job decency (left) and job meaningfulness (right) attributes in the future of work across IT, service-based, and healthcare participants.}
    \label{fig:results_visualization} 
    \Description{Image titled 'How Workers Expected AI to Influence Job Satisfaction'. The top contains a legend with three arrows: upward indicates improvement, level indicates no change, and downward indicates decline. Below are two boxes for 'Decent Work' (left) and 'Meaningful Work' (right), showing how each of the three domains (IT, Service, and Healthcare) is represented for each attribute.}
\end{figure}

Figure~\ref{fig:results_visualization} provides a high-level, comparative overview summarizing how participants across IT, service-based, and healthcare domains imagined AI’s impact on job decency and meaningfulness.
The directional arrows reflect the dominant expectations (in the future of work with AI) expressed by participants within each domain for a given attribute.
We used the upward arrow ($\uparrow$) when the majority of participants within a domain anticipated improvement, or when positive expectations were clearly dominant.
We used the downward arrow ($\downarrow$) when the majority anticipated decline, or when responses were split between no-change and declining satisfaction. We reasoned that this was a decline rather than no change because the latter implies the absence of concern, whereas the presence of an anticipated decline shows a risk to that attribute.
Lastly, we used the hyphen (\textendash{}) when most participants anticipated no change, or when expectations were evenly mixed across improvement, no-change, and decline without a dominant direction.

The visualization (Figure~\ref{fig:results_visualization}) reveals noticeable sectoral differences.
IT workers anticipated improvement across most aspects of both job decency and meaningfulness, while healthcare workers expected more improvement in their job decency compared to meaningfulness. Service workers foresaw almost no change in their job decency, but considerable improvement in their job meaningfulness.

For job decency, IT and healthcare workers commonly expected improvements in working hours, work–life balance, growth opportunities, and working conditions.
By contrast, service workers anticipated no change across most decency attributes except working hours, which they expected to worsen due to the added burden of supervising or correcting AI-driven systems.
Compensation emerged as a distinct point of divergence across domains. Healthcare workers anticipated a decrease in compensation as organizations sought to offset the costs of AI adoption. IT workers, by contrast, expected compensation to increase through expanded growth opportunities enabled by AI. Service workers largely anticipated no change in compensation, due to their tip-based income and existing labor protections.
At the same time, job security was largely expected to remain unchanged across domains, suggesting that participants anticipated shifts in the quality and conditions of work rather than the disappearance of their jobs altogether.
For job meaningfulness, IT workers generally anticipated improvements across multiple aspects, including relationship with colleagues, goal alignment, autonomy, skill variety, and task stimulation.
By contrast, healthcare workers expected minimal change, emphasizing that core clinical responsibility would remain fundamentally human.
However, both anticipated declines in their social image and the credit and recognition they would receive in the workplace. 
Then again, while service workers expected declines in collegial relationships as human coworkers were replaced by AI-driven systems, they also anticipated improvements in autonomy, coworker treatment, and social image, driven by fairer task distribution and the increased respect associated with being perceived as competent in working with advanced technologies.

Taken together, these patterns highlight both opportunities and tensions, showing how AI’s anticipated effects varied across sectors.
For IT workers, who were more familiar with AI technologies, anticipated improvements in decency aspects like workload and growth opportunities tended to reinforce meaningfulness through greater autonomy, skill variety, and alignment with professional goals.
In healthcare, improvements in decent work were largely decoupled from meaningfulness, suggesting that better material conditions might not necessarily enhance the experiential dimensions of work.
In service domain, anticipated declines in decency-related attributes such as working hours coincided with gains in meaningfulness through increased autonomy, better coworker treatment, or enhanced social image, illustrating how AI could simultaneously erode elements of decent work while strengthening aspects of meaningful work.
These cross-dimensional trade-offs show how AI was imagined to reconfigure work not in uniformly positive or negative ways, but through shifting balances between what makes work secure, fair, and materially sustainable, and what makes it engaging, interpersonal, and socially valued.


\section{Discussion}

In this paper, we conducted interviews with participants from three work domains: IT, service-based, and healthcare, to explore the potential effects of introducing AI in their workplaces on their job satisfaction. Drawing on previous research, we asked participants about the decency and meaningfulness of their jobs in their current experiences and how they envision their future work with AI.

Our results highlight the complex and domain-specific nature of AI’s influence on job satisfaction, challenging the often-assumed "one-size-fits-all" notion in human-AI collaboration \cite{verma2023rethinking}.
Although the promise of practical improvements that AI offers, such as task simplification and workload reduction, may appeal to employees in some sectors, others, particularly those in human-facing roles, may view AI as undermining the more affective, meaningful aspects that contribute to their job satisfaction.
Taken together, these results point toward broader implications for how job design and sociotechnical theories of work can be understood or even redefined in AI-mediated workplaces.

Sociotechnical systems theory emphasizes that work is always jointly shaped by social structures (e.g., roles, expectations, norms) and technical systems (e.g., tools, algorithms, workflows), such that changes in one inevitably reconfigure the other \cite{trist1981evolution, baxter2011socio}. From this perspective, the design of technology directly shapes how work roles are defined, how tasks are allocated, and how job design constructs \cite{hackman1976motivation}, such as skill variety, autonomy, recognition, or feedback, are experienced in everyday work.

Building on this perspective, our findings suggest that work mediated by AI introduces a qualitatively different form of role reconfiguration than earlier generations of workplace technologies. Rather than tools being primarily used to optimize workflows \cite{winter2014beyond, abbass2019social, daugherty2019creating}, AI-driven systems increasingly participate in evaluative, decision-making, and representational aspects of work. This participation destabilizes established assumptions about who performs work, who is seen to contribute, and how value is attributed within sociotechnical systems. Consequently, AI is not only perceived as \textit{"a support"} that augments human abilities and reduces workload, but can also emerge as \textit{"a competitor"} in decision-making and in accessing opportunities for visibility and growth. In this sense, AI does not merely alter task allocation but reshapes the social visibility and interpretability of work - dimensions that CSCW research has long shown to be critical for coordination, accountability, and job satisfaction \cite{schmidt2002problem, suchman1995making}.
Furthermore, it is important to note that these dynamics vary across work domains. For example, IT workers often referred to AI as a technical peer whose contributions could be evaluated through shared performance metrics and, in some cases, contribute to higher job satisfaction. In contrast, service and healthcare workers, who emphasized the relational, affective, and moral dimensions of their work, more often perceived AI’s involvement as a threat to important aspects of their professional identity, such as empathy, discretion, and social recognition. This comparison highlights that job satisfaction in AI-mediated work cannot be understood without considering how work is socially evaluated and made visible within a given domain. For CSCW research, this underscores the need to move beyond generic notions of human-AI collaboration and attend more closely to how collaboration is situated within domain-specific norms, values, and accountability structures \cite{sadeghian2025workai}.

All in all, this paper highlights how emerging AI-driven systems not only reshape workflows but also social conditions under which work becomes decent and meaningful. Future research could further explore how alternative design approaches can better enhance job satisfaction in AI-mediated socio-technical systems.

\subsection{Designing for Decency and Meaningfulness}
From a design perspective, our findings suggest that AI-driven systems should not be evaluated solely in terms of efficiency gains or workload reduction. Instead, designers should consider how AI-driven systems mediate visibility, attribution, and opportunities for meaningful engagement in the workplace.
Based on these insights, we highlight five design guidelines for designing AI-driven systems for the future workplace.

\subsubsection{\textbf{Preserve Social Image.}}
Our results showed that AI’s influence on workers’ social image can vary significantly across different sectors.
A recurring concern across domains was the fear that AI would overshadow human contributions, reducing the recognition workers receive for their efforts. Participants anticipated that as AI becomes more capable, it might be perceived as a competitor rather than a collaborator, leading to diminished visibility of human expertise.
For sectors with extensive learning curves, often involving knowledge workers who rely on various forms of knowledge \cite{blackler1995knowledge}, social image is intricately tied to their expertise, efforts required to reach their positions and succeed in their roles, and the subsequent recognition of their contributions. If AI-driven systems dominate visible outcomes, it risks undermining the professional pride of workers in these fields.
In knowledge-intensive domains like IT and healthcare, AI should not dominate and overshadow the worker's expertise but rather provide support \cite{salikutluk2024evaluation}, especially when workers are interacting with clients or patients, to ensure that AI does not appear more competent than the human worker.
This, of course, does not mean that we need to shrink AI’s capabilities to fit those of their human collaborators. Instead, AI-driven systems should be designed so that, while unobtrusively augmenting their human collaborators, they still preserve moments of meaningful interaction and meaningful outcomes at work for them.
In contrast, in jobs similar to service-based roles, collaboration with AI is often viewed as an opportunity to elevate one's social image by demonstrating the ability to work seamlessly with advanced technology, thereby being perceived as more competent and adaptable. Therefore, in such roles, AI-driven systems should be designed in ways that enable workers to integrate them into their work practices and engage with them intuitively while providing them with learning opportunities to enhance their feelings of competence.

\subsubsection{\textbf{Acknowledge the Unseen Human Effort in Human-AI Collaboration.}}
Compensation policies should also recognize and reward the supervisory and corrective tasks performed by employees monitoring AI-driven systems \cite{berezina2019robots}, often referred to as "ghost work" \cite{boeva2023behind}.
The design of AI-driven systems could include mechanisms that clearly delineate the contributions of both human and AI collaborators. For instance, transparent reporting features such as visual dashboards or detailed reports could illustrate the distribution of effort between workers and AI-driven systems, ensuring that human oversight is acknowledged.
Such transparent reporting of contributions resonates with job design principles around task identity and feedback, showing workers how their efforts shape outcomes \cite{hackman1976motivation}.
Furthermore, AI-driven systems should also account for the time and effort employees invest in acquiring new skills or adapting to evolving AI technologies as part of their roles. The training and upskilling required to work effectively with AI should be recognized and valued, ensuring that these efforts are not overlooked, much like the often-undervalued contributions of ghost work.
Such designs could keep humans meaningfully involved in their jobs by enhancing their perceived contributions \cite{xu2024makes} and allowing acknowledgment for their efforts.

\subsubsection{\textbf{Enhance Rather Than Replace Human-Human Interactions.}}
Workplace relationships are a key aspect of job meaningfulness across all sectors.
Existing research also highlights that collaborating with humans is often perceived as more meaningful, as human interactions foster emotional connections and social relatedness \cite{sadeghian2022artificial, ossadnik2023man}. 
While from a negative perspective, AI-driven systems could be a threat to social relationships at work, from a positive perspective, they can enhance opportunities for teamwork through cross-functional collaborations \cite{xiao2025might}, for example, by identifying shared interests or aligning tasks among employees.
In healthcare, AI might encourage cross-departmental collaboration by highlighting mutual availability for patient care discussions. Similarly, in the IT and service sectors, AI could direct workers to seek input from colleagues rather than independently resolving issues, ensuring that social bonds remain intact.
AI should also be designed to enhance aspects of work where interpersonal discussion and decision-making are vital. For instance, service workers value opportunities for customer interaction, which AI should aim to enhance rather than replace. In IT and healthcare, workers often derive satisfaction from collaborative problem-solving, an intellectually stimulating process utilizing a variety of skills that AI should support rather than overshadow.

\subsubsection{\textbf{Balance Autonomy and Control.}}
A standard design recommendation often cited is enabling AI-driven systems to handle repetitive or manual tasks \cite{wangoo2018artificial, liang2024large}, which in theory, could allow employees to focus on higher-level, creative, or decision-making responsibilities \cite{mesko2018will, liang2024large, oh2019physician, socha2020empowering}. However, research has also found that delegating simpler tasks to AI can adversely affect job satisfaction, as it reduces opportunities to engage with less demanding tasks that provide mental relief \cite{waardenburg2024human}. So, the ability to modify the extent of AI involvement would enable workers to focus on tasks that align with their personal goals and expertise. 
However, AI’s influence on workplace autonomy is a double-edged sword. For some, particularly in decision-intensive roles like healthcare, AI’s involvement in tasks such as diagnostics or prescription recommendations might feel like a loss of control, reducing their autonomy.
This could further adversely affect their intellectual stimulation since AI involvement may leave employees with fewer opportunities to exercise their expertise by "stealing their show", reducing their sense of meaningful engagement in their work.
To address this, AI-driven systems could be designed to maintain healthcare professionals' autonomy and authority by allowing them to override or adjust AI recommendations, preserving their decision-making power. For instance, AI could assist in diagnosing or monitoring patients, but while ensuring human oversight \cite{bohr2020rise}.
Conversely, in service-based roles, delegating tasks to AI can enhance feelings of autonomy, as workers perceive themselves as managers directing AI assistants \cite{hemmer2023human}. For instance, wait staff might use AI to manage customer orders or track inventory while focusing on customer interaction, reinforcing their sense of control over their role.

\subsubsection{\textbf{Foster Career Growth.}}
Concerns about technology replacing human labor have existed since the industrial revolution \cite{mokyr2015history, fleming2019robots}.
Such concerns are often tied to perceived reductions in skill variety and growth opportunities \cite{zhan2024there}.
However, in our study, participants did not view AI as threatening their job security; instead, they described AI as supporting skill development, aligning with Herzberg’s motivators in modern, technology-mediated work environments \cite{hertzberg1959motivation}.
In line with sociotechnical perspectives \cite{winter2014beyond}, AI should be designed not to automate expertise away, but to augment employee roles, ensuring their skills evolve and remain relevant in the long term. 
This could be achieved by including feedback loops \cite{bai2022training} in systems, which could positively influence growth opportunities through cross-learning between humans and AI. In doing so, employees can develop new competencies alongside technological advancements, reducing fears of obsolescence, while the AI learns and improves from user interactions \cite{lindvall2021rapid, hafez2021human}.
For example, in IT-domain, AI could coach employees in learning advanced coding through interactive prompts \cite{weisz2025examining}, while in healthcare, it could help medical professionals in refining diagnostic skills by offering insights during complex cases \cite{wang2021brilliant}.

\subsection{Limitations}
While our findings open avenues for future work, they also come with limitations related to the study design and participants.
Although ethnographic futuring offers valuable future-oriented insights, speculating about the future based on current trends and behaviors inevitably involves uncertainty.
In our study, the first half of each interview focused on participants’ current workplace experiences, while the second half prompted them to envision prospective work environments shaped by a self-selected AI.
The answers to these future-oriented questions may have been influenced by participants’ varying levels of AI literacy, information availability, and broader societal (mis)conceptions about AI-driven systems \cite{bradshaw2013seven, cave2019hopes}, often referred to as "imaginaries" in literature \cite{lustig2019intersecting, verma2023rethinking, gomez2025does}.
Nevertheless, this limitation is offset by the value of capturing how individuals make sense of not-yet-existing technologies through the lens of their present-day experiences and perspectives.

Another limitation relates to the composition of our participant sample. While the number of participants was relatively small, this was a deliberate choice aligned with the Interpretative Phenomenological Analysis (IPA) approach, which prioritizes depth and richness of subjective, situated experiences \cite{smith2021interpretative}. 
As such, based on the sample size recommendation for qualitative research by Guest et al. \cite{guest2006many}, we prioritized in-depth engagement with a small number of participants to garner nuanced insights.
Nevertheless, the sample reflects certain imbalances: participants were predominantly male, with no representation of non-binary individuals, and perspectives of older employees were limited, as the oldest participant was 46. Consequently, our findings may underrepresent the perspectives of women and individuals with non-binary gender identities, and may limit insights into how later-career workers anticipate AI’s influence on job decency and meaningfulness. In addition, participants were drawn from different geographical regions (Europe, Asia, and North America), where labor regulations, workplace norms, and societal expectations around AI adoption vary. These characteristics make our findings context-bound and suggest caution when generalizing across genders, age groups, or regions. Future work could address these constraints by engaging larger and more demographically balanced samples, as well as focusing on specific regional contexts to further explore how local labor laws and cultural expectations shape employees’ perceptions of job decency and meaningfulness with AI in the workplace.

\section{Conclusion}
In this paper, we explored how workers across IT, service, and healthcare domains imagine AI to reshape job decency and meaningfulness — two foundational dimensions of job satisfaction. Our findings show that AI’s anticipated impact is neither uniform nor universally positive - while some aspects of work may be enhanced, others risk being threatened in ways that differ sharply across domains.
By analyzing job satisfaction through the dual lens of decency and meaningfulness, this paper highlights where AI is expected to support workers (e.g., enhanced skill variety or better work-life balance) and where it may create new tensions (e.g., threatening recognition or altering social image). Taken together, our results challenge assumptions that AI will influence all workers equally and underscore the need for domain-sensitive, value-aligned systems.
These insights provide guidance for designing AI-driven systems for the future of work that balance job decency and meaningfulness, so that AI may augment, rather than diminish, job satisfaction.

\section{Acknowledgments}
This research was conducted as part of the Sensing \& Sensibility research project at the University of Siegen (\url{https://sensing.uni-siegen.de/}). We thank the participants and colleagues whose support and valuable input contributed to this work.


\bibliographystyle{ACM-Reference-Format}
\bibliography{base}

@String{Computing = "Computing" }

@String{Computer = "{IEEE} Computer" }

@String{Springer = "Springer-Verlag" }

@book{office2014global,
  title={Global Employment Trends 2014: Risk of a Jobless Recovery?},
  author={Office, I.L.},
  isbn={9789221274858},
  series={Global employment trends},
  url={https://books.google.de/books?id=_iK5oAEACAAJ},
  year={2014},
  publisher={International Labour Organisation (ILO)}
}

@article{duffy2016psychology,
  title={The psychology of working theory.},
  author={Duffy, Ryan D and Blustein, David L and Diemer, Matthew A and Autin, Kelsey L},
  journal={Journal of counseling psychology},
  volume={63},
  number={2},
  pages={127},
  year={2016},
  publisher={American Psychological Association}
}

@article{blustein2023understanding,
  title={Understanding decent work and meaningful work},
  author={Blustein, David L and Lysova, Evgenia I and Duffy, Ryan D},
  journal={Annual Review of Organizational Psychology and Organizational Behavior},
  volume={10},
  pages={289--314},
  year={2023},
  publisher={Annual Reviews}
}

@article{allan2020decent,
  title={Decent and meaningful work: A longitudinal study.},
  author={Allan, Blake A and Autin, Kelsey L and Duffy, Ryan D and Sterling, Haley M},
  journal={Journal of Counseling Psychology},
  volume={67},
  number={6},
  pages={669},
  year={2020},
  publisher={American Psychological Association}
}

@article{rosso2010meaning,
  title={On the meaning of work: A theoretical integration and review},
  author={Rosso, Brent D and Dekas, Kathryn H and Wrzesniewski, Amy},
  journal={Research in organizational behavior},
  volume={30},
  pages={91--127},
  year={2010},
  publisher={Elsevier}
}

@article{steger2012measuring,
  title={Measuring meaningful work: The work and meaning inventory (WAMI)},
  author={Steger, Michael F and Dik, Bryan J and Duffy, Ryan D},
  journal={Journal of career Assessment},
  volume={20},
  number={3},
  pages={322--337},
  year={2012},
  publisher={Sage Publications Sage CA: Los Angeles, CA}
}

@article{dik2009calling,
  title={Calling and vocation at work: Definitions and prospects for research and practice},
  author={Dik, Bryan J and Duffy, Ryan D},
  journal={The counseling psychologist},
  volume={37},
  number={3},
  pages={424--450},
  year={2009},
  publisher={Sage Publications Sage CA: Los Angeles, CA}
}

@article{lysova2019meaningful,
  title={Meaningful Work and Family},
  author={Lysova, Evgenia I},
  journal={The Oxford Handbook of Meaningful Work},
  pages={404},
  year={2019},
  publisher={Oxford University Press}
}

@article{pratt2003fostering,
  title={Fostering meaningfulness in working and at work},
  author={Pratt, Michael G and Ashforth, Blake E},
  journal={Positive organizational scholarship: Foundations of a new discipline},
  volume={309},
  pages={327},
  year={2003}
}

@article{aguinis2019corporate,
  title={On corporate social responsibility, sensemaking, and the search for meaningfulness through work},
  author={Aguinis, Herman and Glavas, Ante},
  journal={Journal of management},
  volume={45},
  number={3},
  pages={1057--1086},
  year={2019},
  publisher={Sage Publications Sage CA: Los Angeles, CA}
}

@article{fletcher2019can,
  title={How can personal development lead to increased engagement? The roles of meaningfulness and perceived line manager relations},
  author={Fletcher, Luke},
  journal={The International Journal of Human Resource Management},
  volume={30},
  number={7},
  pages={1203--1226},
  year={2019},
  publisher={Taylor \& Francis}
}

@article{wrzesniewski2003interpersonal,
  title={Interpersonal sensemaking and the meaning of work},
  author={Wrzesniewski, Amy and Dutton, Jane E and Debebe, Gelaye},
  journal={Research in organizational behavior},
  volume={25},
  pages={93--135},
  year={2003},
  publisher={Elsevier}
}

@article{colbert2016flourishing,
  title={Flourishing via workplace relationships: Moving beyond instrumental support},
  author={Colbert, Amy E and Bono, Joyce E and Purvanova, Radostina K},
  journal={Academy of Management Journal},
  volume={59},
  number={4},
  pages={1199--1223},
  year={2016},
  publisher={Academy of Management Briarcliff Manor, NY}
}

@article{carton2018m,
  title={“I’m not mopping the floors, I’m putting a man on the moon”: How NASA leaders enhanced the meaningfulness of work by changing the meaning of work},
  author={Carton, Andrew M},
  journal={Administrative Science Quarterly},
  volume={63},
  number={2},
  pages={323--369},
  year={2018},
  publisher={Sage Publications Sage CA: Los Angeles, CA}
}

@article{hackman1976motivation,
  title={Motivation through the design of work: Test of a theory},
  author={Hackman, J Richard and Oldham, Greg R},
  journal={Organizational behavior and human performance},
  volume={16},
  number={2},
  pages={250--279},
  year={1976},
  publisher={Elsevier}
}

@article{kim2020thriving,
  title={Thriving on demand: Challenging work results in employee flourishing through appraisals and resources.},
  author={Kim, Minseo and Beehr, Terry A},
  journal={International Journal of Stress Management},
  volume={27},
  number={2},
  pages={111},
  year={2020},
  publisher={Educational Publishing Foundation}
}

@article{grant2007relational,
  title={Relational job design and the motivation to make a prosocial difference},
  author={Grant, Adam M},
  journal={Academy of management review},
  volume={32},
  number={2},
  pages={393--417},
  year={2007},
  publisher={Academy of Management Briarcliff Manor, NY 10510}
}

@article{grant2007impact,
  title={Impact and the art of motivation maintenance: The effects of contact with beneficiaries on persistence behavior},
  author={Grant, Adam M and Campbell, Elizabeth M and Chen, Grace and Cottone, Keenan and Lapedis, David and Lee, Karen},
  journal={Organizational behavior and human decision processes},
  volume={103},
  number={1},
  pages={53--67},
  year={2007},
  publisher={Elsevier}
}

@article{arnoux2016perceived,
  title={Perceived work conditions and turnover intentions: The mediating role of meaning of work},
  author={Arnoux-Nicolas, Caroline and Sovet, Laurent and Lhotellier, Lin and Di Fabio, Annamaria and Bernaud, Jean-Luc},
  journal={Frontiers in psychology},
  volume={7},
  pages={704},
  year={2016},
  publisher={Frontiers Media SA}
}

@article{lips2020effect,
  title={The effect of fairness, responsible leadership and worthy work on multiple dimensions of meaningful work},
  author={Lips-Wiersma, Marjolein and Haar, Jarrod and Wright, Sarah},
  journal={Journal of business ethics},
  volume={161},
  pages={35--52},
  year={2020},
  publisher={Springer}
}

@book{lazar2017research,
  title={Research methods in human-computer interaction},
  author={Lazar, Jonathan and Feng, Jinjuan Heidi and Hochheiser, Harry},
  year={2017},
  publisher={Morgan Kaufmann}
}

@article{pereira2019empirical,
  title={Empirical research on decent work: A literature review},
  author={Pereira, Susana and dos Santos, Nuno Rebelo and Pais, Leonor},
  year={2019},
  publisher={Scandinavian Journal of Work and Organizational Psychology}
}

@article{sadeghian2024soul,
  title={The Soul of Work: Evaluation of Job Meaningfulness and Accountability in Human-AI Collaboration},
  author={Sadeghian, Shadan and Uhde, Alarith and Hassenzahl, Marc},
  journal={Proceedings of the ACM on Human-Computer Interaction},
  volume={8},
  number={CSCW1},
  pages={1--26},
  year={2024},
  publisher={ACM New York, NY, USA}
}

@article{ahmad2013paradigms,
  title={Paradigms of quality of work life},
  author={Ahmad, Shoeb},
  journal={Journal of Human Values},
  volume={19},
  number={1},
  pages={73--82},
  year={2013},
  publisher={Sage Publications Sage India: New Delhi, India}
}

@article{marquis2024proliferation,
  title={Proliferation of AI tools: A multifaceted evaluation of user perceptions and emerging trend},
  author={Marquis, Yewande and Oladoyinbo, Tunboson Oyewale and Olabanji, Samuel Oladiipo and Olaniyi, Oluwaseun Oladeji and Ajayi, Samson Abidemi},
  journal={Asian Journal of Advanced Research and Reports},
  volume={18},
  number={1},
  pages={30--55},
  year={2024}
}

@article{nyholm2020can,
  title={Can a robot be a good colleague?},
  author={Nyholm, Sven and Smids, Jilles},
  journal={Science and engineering ethics},
  volume={26},
  pages={2169--2188},
  year={2020},
  publisher={Springer}
}

@inproceedings{sadeghian2022artificial,
  title={The” artificial” colleague: evaluation of work satisfaction in collaboration with non-human coworkers},
  author={Sadeghian, Shadan and Hassenzahl, Marc},
  booktitle={27th International Conference on Intelligent User Interfaces},
  pages={27--35},
  year={2022}
}

@inproceedings{lindley2014anticipatory,
  title={Anticipatory Ethnography: Design fiction as an input to design ethnography},
  author={Lindley, Joseph and Sharma, Dhruv and Potts, Robert},
  booktitle={Ethnographic Praxis in Industry Conference Proceedings},
  volume={2014},
  number={1},
  pages={237--253},
  year={2014},
  organization={Wiley Online Library}
}

@article{dwivedi2021artificial,
  title={Artificial Intelligence (AI): Multidisciplinary perspectives on emerging challenges, opportunities, and agenda for research, practice and policy},
  author={Dwivedi, Yogesh K and Hughes, Laurie and Ismagilova, Elvira and Aarts, Gert and Coombs, Crispin and Crick, Tom and Duan, Yanqing and Dwivedi, Rohita and Edwards, John and Eirug, Aled and others},
  journal={International Journal of Information Management},
  volume={57},
  pages={101994},
  year={2021},
  publisher={Elsevier}
}

@article{miller2018ai,
  title={AI: Augmentation, more so than automation},
  author={Miller, Steven M},
  year={2018},
  publisher={Singapore Management University, Centre for Management Practice}
}

@book{daugherty2018human+,
  title={Human+ machine: Reimagining work in the age of AI},
  author={Daugherty, Paul R and Wilson, H James},
  year={2018},
  publisher={Harvard Business Press}
}

@article{gowan2014moving,
  title={Moving from job loss to career management: The past, present, and future of involuntary job loss research},
  author={Gowan, Mary A},
  journal={Human Resource Management Review},
  volume={24},
  number={3},
  pages={258--270},
  year={2014},
  publisher={Elsevier}
}

@article{wanberg2012individual,
  title={The individual experience of unemployment},
  author={Wanberg, Connie R},
  journal={Annual review of psychology},
  volume={63},
  pages={369--396},
  year={2012},
  publisher={Annual Reviews}
}

@article{paul2009unemployment,
  title={Unemployment impairs mental health: Meta-analyses},
  author={Paul, Karsten I and Moser, Klaus},
  journal={Journal of Vocational behavior},
  volume={74},
  number={3},
  pages={264--282},
  year={2009},
  publisher={Elsevier}
}

@article{allen2012work,
  title={The work--family role interface: a synthesis of the research from industrial and organizational psychology},
  author={Allen, Tammy D},
  journal={Handbook of Psychology, Second Edition},
  volume={12},
  year={2012},
  publisher={Wiley Online Library}
}

@article{greenhaus2011work,
  title={Work--family balance: A review and extension of the literature.},
  author={Greenhaus, Jeffrey H and Allen, Tammy D},
  year={2011},
  publisher={American Psychological Association}
}

@article{greenhaus2014contemporary,
  title={The contemporary career: A work--home perspective},
  author={Greenhaus, Jeffrey H and Kossek, Ellen Ernst},
  journal={Annu. Rev. Organ. Psychol. Organ. Behav.},
  volume={1},
  number={1},
  pages={361--388},
  year={2014},
  publisher={Annual Reviews}
}

@article{sullivan2009advances,
  title={Advances in career theory and research: A critical review and agenda for future exploration},
  author={Sullivan, Sherry E and Baruch, Yehuda},
  journal={Journal of management},
  volume={35},
  number={6},
  pages={1542--1571},
  year={2009},
  publisher={Sage Publications Sage CA: Los Angeles, CA}
}

@article{baruch2006career,
  title={Career development in organizations and beyond: Balancing traditional and contemporary viewpoints},
  author={Baruch, Yehuda},
  journal={Human resource management review},
  volume={16},
  number={2},
  pages={125--138},
  year={2006},
  publisher={Elsevier}
}

@article{judge2010relationship,
  title={The relationship between pay and job satisfaction: A meta-analysis of the literature},
  author={Judge, Timothy A and Piccolo, Ronald F and Podsakoff, Nathan P and Shaw, John C and Rich, Bruce L},
  journal={Journal of vocational behavior},
  volume={77},
  number={2},
  pages={157--167},
  year={2010},
  publisher={Elsevier}
}

@article{dulebohn2007compensation,
  title={Compensation research past, present, and future},
  author={Dulebohn, James H and Werling, Stephen E},
  journal={Human Resource Management Review},
  volume={17},
  number={2},
  pages={191--207},
  year={2007},
  publisher={Elsevier}
}

@article{fouche2017antecedents,
  title={Antecedents and outcomes of meaningful work among school teachers},
  author={Fouch{\'e}, Elmari and Rothmann, Sebastiaan Snr and Van der Vyver, Corne},
  journal={SA Journal of Industrial Psychology},
  volume={43},
  number={1},
  pages={1--10},
  year={2017},
  publisher={AOSIS Publishing}
}

@article{lysova2019fostering,
  title={Fostering meaningful work in organizations: A multi-level review and integration},
  author={Lysova, Evgenia I and Allan, Blake A and Dik, Bryan J and Duffy, Ryan D and Steger, Michael F},
  journal={Journal of vocational behavior},
  volume={110},
  pages={374--389},
  year={2019},
  publisher={Elsevier}
}

@article{bailey2016makes,
  title={What makes work meaningful—or meaningless},
  author={Bailey, Catherine and Madden, Adrian},
  journal={MIT Sloan management review},
  year={2016}
}

@article{martela2018autonomy,
  title={Autonomy, competence, relatedness, and beneficence: A multicultural comparison of the four pathways to meaningful work},
  author={Martela, Frank and Riekki, Tapani JJ},
  journal={Frontiers in psychology},
  volume={9},
  pages={327587},
  year={2018},
  publisher={Frontiers}
}

@book{russell2016artificial,
  title={Artificial intelligence: a modern approach},
  author={Russell, Stuart J and Norvig, Peter},
  year={2016},
  publisher={Pearson}
}

@book{miller2017robots,
  title={Robots and robotics: principles, systems, and industrial applications},
  author={Miller, Mark R and Miller, Rex},
  year={2017},
  publisher={McGraw-Hill Education}
}

@book{bhaumik2018ai,
  title={From AI to robotics: mobile, social, and sentient robots},
  author={Bhaumik, Arkapravo},
  year={2018},
  publisher={CrC Press}
}

@article{bhargava2021employees,
  title={Employees’ perceptions of the implementation of robotics, artificial intelligence, and automation (RAIA) on job satisfaction, job security, and employability},
  author={Bhargava, Amisha and Bester, Marais and Bolton, Lucy},
  journal={Journal of Technology in Behavioral Science},
  volume={6},
  number={1},
  pages={106--113},
  year={2021},
  publisher={Springer}
}

@article{nam2019technology,
  title={Technology usage, expected job sustainability, and perceived job insecurity},
  author={Nam, Taewoo},
  journal={Technological Forecasting and Social Change},
  volume={138},
  pages={155--165},
  year={2019},
  publisher={Elsevier}
}

@article{findlay2017employer,
  title={Employer choice and job quality: Workplace innovation, work redesign, and employee perceptions of job quality in a complex health-care setting},
  author={Findlay, Patricia and Lindsay, Colin and McQuarrie, Jo and Bennie, Marion and Corcoran, Emma D and Van Der Meer, Robert},
  journal={Work and Occupations},
  volume={44},
  number={1},
  pages={113--136},
  year={2017},
  publisher={Sage Publications Sage CA: Los Angeles, CA}
}

@article{smids2020robots,
  title={Robots in the workplace: a threat to—or opportunity for—meaningful work?},
  author={Smids, Jilles and Nyholm, Sven and Berkers, Hannah},
  journal={Philosophy \& Technology},
  volume={33},
  number={3},
  pages={503--522},
  year={2020},
  publisher={Springer}
}

@inproceedings{hemmer2023human,
  title={Human-AI Collaboration: The Effect of AI Delegation on Human Task Performance and Task Satisfaction},
  author={Hemmer, Patrick and Westphal, Monika and Schemmer, Max and Vetter, Sebastian and V{\"o}ssing, Michael and Satzger, Gerhard},
  booktitle={Proceedings of the 28th International Conference on Intelligent User Interfaces},
  pages={453--463},
  year={2023}
}

@inproceedings{hirst2014does,
  title={Does technological innovation increase unemployment},
  author={Hirst, T},
  booktitle={The World Economic Forum Blog [online], Agenda Retrieved (https://agenda. weforum. org/2014/11/does-technologicalinnovationincreaseunemployment},
  year={2014}
}

@misc{agrawal2017expect,
  title={What to expect from artificial intelligence},
  author={Agrawal, Ajay and Gans, Joshua and Goldfarb, Avi},
  year={2017},
  publisher={MIT Sloan Management Review Cambridge, MA, USA}
}

@article{ivanov2017robonomics,
  title={Robonomics-principles, benefits, challenges, solutions},
  author={Ivanov, Stanislav Hristov},
  year={2017}
}

@article{davenport2018artificial,
  title={Artificial intelligence for the real world},
  author={Davenport, Thomas H and Ronanki, Rajeev and others},
  journal={Harvard business review},
  volume={96},
  number={1},
  pages={108--116},
  year={2018}
}

@article{seppala2013social,
  title={Social connection and compassion: Important predictors of health and well-being},
  author={Seppala, Emma and Rossomando, Timothy and Doty, James R},
  journal={Social Research: An International Quarterly},
  volume={80},
  number={2},
  pages={411--430},
  year={2013},
  publisher={Johns Hopkins University Press}
}

@article{chui2015four,
  title={Four fundamentals of workplace automation},
  author={Chui, Michael and Manyika, James and Miremadi, Mehdi},
  journal={McKinsey Quarterly},
  volume={29},
  number={3},
  pages={1--9},
  year={2015}
}

@article{kolbjornsrud2016promise,
  title={The promise of artificial intelligence},
  author={Kolbj{\o}rnsrud, Vegard and Amico, Richard and Thomas, Robert J},
  journal={Accenture: Dublin, Ireland},
  year={2016}
}

@article{mcclure2018you,
  title={“You’re fired,” says the robot: The rise of automation in the workplace, technophobes, and fears of unemployment},
  author={McClure, Paul K},
  journal={Social Science Computer Review},
  volume={36},
  number={2},
  pages={139--156},
  year={2018},
  publisher={Sage Publications Sage CA: Los Angeles, CA}
}

@article{parasuraman2008situation,
  title={Situation awareness, mental workload, and trust in automation: Viable, empirically supported cognitive engineering constructs},
  author={Parasuraman, Raja and Sheridan, Thomas B and Wickens, Christopher D},
  journal={Journal of cognitive engineering and decision making},
  volume={2},
  number={2},
  pages={140--160},
  year={2008},
  publisher={Sage Publications Sage CA: Los Angeles, CA}
}

@article{sparks2001explaining,
  title={Explaining the effects of transformational leadership: an investigation of the effects of higher-order motives in multilevel marketing organizations},
  author={Sparks, John R and Schenk, Joseph A},
  journal={Journal of Organizational Behavior: The International Journal of Industrial, Occupational and Organizational Psychology and Behavior},
  volume={22},
  number={8},
  pages={849--869},
  year={2001},
  publisher={Wiley Online Library}
}

@article{lips2009discriminating,
  title={Discriminating between ‘meaningful work’and the ‘management of meaning’},
  author={Lips-Wiersma, Marjolein and Morris, Lani},
  journal={Journal of business ethics},
  volume={88},
  pages={491--511},
  year={2009},
  publisher={Springer}
}

@article{laschke2020positive,
  title={Positive work practices. Opportunities and challenges in designing meaningful work-related technology},
  author={Laschke, Matthias and Uhde, Alarith and Hassenzahl, Marc},
  journal={arXiv preprint arXiv:2003.05533},
  year={2020}
}

@book{terkel1974working,
  title={Working: People talk about what they do all day and how they feel about what they do},
  author={Terkel, Studs},
  year={1974},
  publisher={The New Press}
}

@inproceedings{baldauf2021automation,
  title={Automation experience at the workplace},
  author={Baldauf, Matthias and Fr{\"o}hlich, Peter and Sadeghian, Shadan and Palanque, Philippe and Roto, Virpi and Ju, Wendy and Baillie, Lynne and Tscheligi, Manfred},
  booktitle={Extended Abstracts of the 2021 CHI Conference on Human Factors in Computing Systems},
  pages={1--6},
  year={2021}
}

@article{locke1969job,
  title={What is job satisfaction?},
  author={Locke, Edwin A},
  journal={Organizational behavior and human performance},
  volume={4},
  number={4},
  pages={309--336},
  year={1969},
  publisher={Elsevier}
}

@article{henne1985job,
  title={Job dissatisfaction: What are the consequences?},
  author={Henne, Douglas and Locke, Edwin A},
  journal={International journal of psychology},
  volume={20},
  number={2},
  pages={221--240},
  year={1985},
  publisher={Wiley Online Library}
}

@article{zopiatis2014job,
  title={Job involvement, commitment, satisfaction and turnover: Evidence from hotel employees in Cyprus},
  author={Zopiatis, Anastasios and Constanti, Panayiotis and Theocharous, Antonis L},
  journal={Tourism Management},
  volume={41},
  pages={129--140},
  year={2014},
  publisher={Elsevier}
}

@article{kong2018job,
  title={Job satisfaction research in the field of hospitality and tourism},
  author={Kong, Haiyan and Jiang, Xinyu and Chan, Wilco and Zhou, Xiaoge},
  journal={International journal of contemporary hospitality management},
  volume={30},
  number={5},
  pages={2178--2194},
  year={2018},
  publisher={Emerald Publishing Limited}
}

@article{dorta2021effects,
  title={Effects of high-performance work systems (HPWS) on hospitality employees’ outcomes through their organizational commitment, motivation, and job satisfaction},
  author={Dorta-Afonso, Daniel and Gonz{\'a}lez-de-la-Rosa, Manuel and Garcia-Rodriguez, Francisco J and Romero-Dom{\'\i}nguez, Laura},
  journal={Sustainability},
  volume={13},
  number={6},
  pages={3226},
  year={2021},
  publisher={MDPI}
}

@article{ossadnik2023man,
  title={Man or machine--or something in between? Social responses to voice assistants at work and their effects on job satisfaction},
  author={Ossadnik, Jonas and Muehlfeld, Katrin and Goerke, Laszlo},
  journal={Computers in Human Behavior},
  volume={149},
  pages={107919},
  year={2023},
  publisher={Elsevier}
}

@article{walliser2019team,
  title={Team structure and team building improve human--machine teaming with autonomous agents},
  author={Walliser, James C and de Visser, Ewart J and Wiese, Eva and Shaw, Tyler H},
  journal={Journal of Cognitive Engineering and Decision Making},
  volume={13},
  number={4},
  pages={258--278},
  year={2019},
  publisher={SAGE Publications Sage CA: Los Angeles, CA}
}

@article{bordot2022artificial,
  title={Artificial intelligence, robots and unemployment: evidence from OECD countries},
  author={Bordot, Florent},
  journal={Journal of innovation economics \& management},
  number={1},
  pages={117--138},
  year={2022},
  publisher={Cairn/Softwin}
}

@article{crandall2018cooperating,
  title={Cooperating with machines},
  author={Crandall, Jacob W and Oudah, Mayada and Tennom and Ishowo-Oloko, Fatimah and Abdallah, Sherief and Bonnefon, Jean-Fran{\c{c}}ois and Cebrian, Manuel and Shariff, Azim and Goodrich, Michael A and Rahwan, Iyad},
  journal={Nature communications},
  volume={9},
  number={1},
  pages={233},
  year={2018},
  publisher={Nature Publishing Group UK London}
}

@article{akata2020research,
  title={A research agenda for hybrid intelligence: augmenting human intellect with collaborative, adaptive, responsible, and explainable artificial intelligence},
  author={Akata, Zeynep and Balliet, Dan and De Rijke, Maarten and Dignum, Frank and Dignum, Virginia and Eiben, Guszti and Fokkens, Antske and Grossi, Davide and Hindriks, Koen and Hoos, Holger and others},
  journal={Computer},
  volume={53},
  number={8},
  pages={18--28},
  year={2020},
  publisher={IEEE}
}

@article{de2021ai,
  title={AI should augment human intelligence, not replace it},
  author={De Cremer, David and Kasparov, Garry},
  journal={Harvard Business Review},
  volume={18},
  number={1},
  year={2021}
}

@article{dellermann2021future,
  title={The future of human-AI collaboration: a taxonomy of design knowledge for hybrid intelligence systems},
  author={Dellermann, Dominik and Calma, Adrian and Lipusch, Nikolaus and Weber, Thorsten and Weigel, Sascha and Ebel, Philipp},
  journal={arXiv preprint arXiv:2105.03354},
  year={2021}
}

@article{braga2017emperor,
  title={The emperor of strong AI has no clothes: limits to artificial intelligence},
  author={Braga, Adriana and Logan, Robert K},
  journal={Information},
  volume={8},
  number={4},
  pages={156},
  year={2017},
  publisher={MDPI}
}

@inproceedings{esterwood2020human,
  title={Human robot team design},
  author={Esterwood, Connor and Robert, Lionel P},
  booktitle={Proceedings of the 8th international conference on human-agent interaction},
  pages={251--253},
  year={2020}
}

@article{you2018enhancing,
  title={Enhancing perceived safety in human--robot collaborative construction using immersive virtual environments},
  author={You, Sangseok and Kim, Jeong-Hwan and Lee, SangHyun and Kamat, Vineet and Robert Jr, Lionel P},
  journal={Automation in Construction},
  volume={96},
  pages={161--170},
  year={2018},
  publisher={Elsevier}
}

@article{raftopoulos2023human,
  title={Human-AI collaboration in organisations: A literature review on enabling value creation},
  author={Raftopoulos, Marigo and Hamari, Juho},
  year={2023}
}

@inproceedings{qian2024take,
  title={Take It, Leave It, or Fix It: Measuring Productivity and Trust in Human-AI Collaboration},
  author={Qian, Crystal and Wexler, James},
  booktitle={Proceedings of the 29th International Conference on Intelligent User Interfaces},
  pages={370--384},
  year={2024}
}

@inproceedings{lindvall2021rapid,
  title={Rapid assisted visual search: Supporting digital pathologists with imperfect AI},
  author={Lindvall, Martin and Lundstr{\"o}m, Claes and L{\"o}wgren, Jonas},
  booktitle={Proceedings of the 26th International Conference on Intelligent User Interfaces},
  pages={504--513},
  year={2021}
}

@inproceedings{han2024teams,
  title={When Teams Embrace AI: Human Collaboration Strategies in Generative Prompting in a Creative Design Task},
  author={Han, Yuanning and Qiu, Ziyi and Cheng, Jiale and LC, RAY},
  booktitle={Proceedings of the CHI Conference on Human Factors in Computing Systems},
  pages={1--14},
  year={2024}
}

@inproceedings{xu2023comparing,
  title={Comparing zealous and restrained ai recommendations in a real-world human-ai collaboration task},
  author={Xu, Chengyuan and Lien, Kuo-Chin and H{\"o}llerer, Tobias},
  booktitle={Proceedings of the 2023 CHI Conference on Human Factors in Computing Systems},
  pages={1--15},
  year={2023}
}

@inproceedings{xu2024makes,
  title={What Makes It Mine? Exploring Psychological Ownership over Human-AI Co-Creations},
  author={Xu, Yuxin and Cheng, Mengqiu and Kuzminykh, Anastasia},
  booktitle={Graphics Interface 2024 Second Deadline},
  year={2024}
}

@inproceedings{kobiella2024if,
  title={" If the Machine Is As Good As Me, Then What Use Am I?"--How the Use of ChatGPT Changes Young Professionals' Perception of Productivity and Accomplishment},
  author={Kobiella, Charlotte and Flores L{\'o}pez, Yarhy Said and Waltenberger, Franz and Draxler, Fiona and Schmidt, Albrecht},
  booktitle={Proceedings of the CHI Conference on Human Factors in Computing Systems},
  pages={1--16},
  year={2024}
}

@inproceedings{guo2024exploring,
  title={Exploring the Impact of AI Value Alignment in Collaborative Ideation: Effects on Perception, Ownership, and Output},
  author={Guo, Alicia and Pataranutaporn, Pat and Maes, Pattie},
  booktitle={Extended Abstracts of the CHI Conference on Human Factors in Computing Systems},
  pages={1--11},
  year={2024}
}

@inproceedings{gu2024data,
  title={How do data analysts respond to ai assistance? a wizard-of-oz study},
  author={Gu, Ken and Grunde-McLaughlin, Madeleine and McNutt, Andrew and Heer, Jeffrey and Althoff, Tim},
  booktitle={Proceedings of the CHI Conference on Human Factors in Computing Systems},
  pages={1--22},
  year={2024}
}

@article{waardenburg2024human,
  title={Human-AI Collaboration: A Blessing or a Curse for Safety at Work?},
  author={Waardenburg, Lauren},
  journal={Tecnoscienza--Italian Journal of Science \& Technology Studies},
  volume={15},
  number={1},
  pages={133--146},
  year={2024}
}

@inproceedings{wangoo2018artificial,
  title={Artificial intelligence techniques in software engineering for automated software reuse and design},
  author={Wangoo, Divanshi Priyadarshni},
  booktitle={2018 4th International conference on computing communication and automation (ICCCA)},
  pages={1--4},
  year={2018},
  organization={IEEE}
}

@inproceedings{liang2024large,
  title={A large-scale survey on the usability of ai programming assistants: Successes and challenges},
  author={Liang, Jenny T and Yang, Chenyang and Myers, Brad A},
  booktitle={Proceedings of the 46th IEEE/ACM International Conference on Software Engineering},
  pages={1--13},
  year={2024}
}

@incollection{bohr2020rise,
  title={The rise of artificial intelligence in healthcare applications},
  author={Bohr, Adam and Memarzadeh, Kaveh},
  booktitle={Artificial Intelligence in healthcare},
  pages={25--60},
  year={2020},
  publisher={Elsevier}
}

@article{socha2020empowering,
  title={Empowering the health workforce: Strategies to make the most of the digital revolution},
  author={Socha-Dietrich, Karolina},
  journal={Organisation for Economic Co-operation and Development (OECD)},
  year={2020}
}

@article{oh2019physician,
  title={Physician confidence in artificial intelligence: an online mobile survey},
  author={Oh, Songhee and Kim, Jae Heon and Choi, Sung-Woo and Lee, Hee Jeong and Hong, Jungrak and Kwon, Soon Hyo},
  journal={Journal of medical Internet research},
  volume={21},
  number={3},
  pages={e12422},
  year={2019},
  publisher={JMIR Publications Toronto, Canada}
}

@article{mesko2018will,
  title={Will artificial intelligence solve the human resource crisis in healthcare?},
  author={Mesk{\'o}, Bertalan and Het{\'e}nyi, Gergely and Gy{\H{o}}rffy, Zsuzsanna},
  journal={BMC health services research},
  volume={18},
  pages={1--4},
  year={2018},
  publisher={Springer}
}

@inproceedings{boeva2023behind,
  title={Behind the Scenes of Automation: Ghostly Care-Work, Maintenance, and Interferences: Exploring participatory practices and methods to uncover the ghostly presence of humans and human labor in automation},
  author={Boeva, Yana and Berger, Arne and Bischof, Andreas and Doggett, Olivia and Heuer, Hendrik and Jarke, Juliane and Treusch, Pat and S{\o}raa, Roger Andre and Tacheva, Zhasmina and Voigt, Maja-Lee},
  booktitle={Extended Abstracts of the 2023 CHI Conference on Human Factors in Computing Systems},
  pages={1--5},
  year={2023}
}

@incollection{berezina2019robots,
  title={Robots, artificial intelligence, and service automation in restaurants},
  author={Berezina, Katerina and Ciftci, Olena and Cobanoglu, Cihan},
  booktitle={Robots, artificial intelligence, and service automation in travel, tourism and hospitality},
  pages={185--219},
  year={2019},
  publisher={Emerald Publishing Limited}
}

@inproceedings{hafez2021human,
  title={Human digital twin—enabling human-agents collaboration},
  author={Hafez, Wael},
  booktitle={2021 4th International Conference on Intelligent Robotics and Control Engineering (IRCE)},
  pages={40--45},
  year={2021},
  organization={IEEE}
}

@article{blackler1995knowledge,
  title={Knowledge, knowledge work and organizations: An overview and interpretation},
  author={Blackler, Frank},
  journal={Organization studies},
  volume={16},
  number={6},
  pages={1021--1046},
  year={1995},
  publisher={Sage Publications Sage CA: Thousand Oaks, CA}
}

@inproceedings{salikutluk2024evaluation,
  title={An Evaluation of Situational Autonomy for Human-AI Collaboration in a Shared Workspace Setting},
  author={Salikutluk, Vildan and Sch{\"o}pper, Janik and Herbert, Franziska and Scheuermann, Katrin and Frodl, Eric and Balfanz, Dirk and J{\"a}kel, Frank and Koert, Dorothea},
  booktitle={Proceedings of the CHI Conference on Human Factors in Computing Systems},
  pages={1--17},
  year={2024}
}

@article{sigov2024emerging,
  title={Emerging enabling technologies for industry 4.0 and beyond},
  author={Sigov, Alexander and Ratkin, Leonid and Ivanov, Leonid A and Xu, Li Da},
  journal={Information Systems Frontiers},
  volume={26},
  number={5},
  pages={1585--1595},
  year={2024},
  publisher={Springer}
}

@article{limna2023artificial,
  title={Artificial Intelligence (AI) in the hospitality industry: A review article},
  author={Limna, Pongsakorn},
  journal={International Journal of Computing Sciences Research},
  volume={7},
  pages={1306--1317},
  year={2023}
}

@article{shaheen2021applications,
  title={Applications of Artificial Intelligence (AI) in healthcare: A review},
  author={Shaheen, Mohammed Yousef},
  journal={ScienceOpen Preprints},
  year={2021},
  publisher={ScienceOpen}
}

@article{hertzberg1959motivation,
  title={The motivation to work},
  author={Hertzberg, Frederick and Mausner, Bernard and Snyderman, Barbara},
  journal={New York},
  year={1959}
}

@article{bradshaw2013seven,
  title={The seven deadly myths of" autonomous systems"},
  author={Bradshaw, Jeffrey M and Hoffman, Robert R and Woods, David D and Johnson, Matthew},
  journal={IEEE Intelligent Systems},
  volume={28},
  number={3},
  pages={54--61},
  year={2013},
  publisher={IEEE}
}

@article{cave2019hopes,
  title={Hopes and fears for intelligent machines in fiction and reality},
  author={Cave, Stephen and Dihal, Kanta},
  journal={Nature machine intelligence},
  volume={1},
  number={2},
  pages={74--78},
  year={2019},
  publisher={Nature Publishing Group UK London}
}

@article{lustig2019intersecting,
  title={Intersecting imaginaries: visions of decentralized autonomous systems},
  author={Lustig, Caitlin},
  journal={Proceedings of the ACM on Human-Computer Interaction},
  volume={3},
  number={CSCW},
  pages={1--27},
  year={2019},
  publisher={ACM New York, NY, USA}
}

@inproceedings{verma2023rethinking,
  title={Rethinking the role of AI with physicians in oncology: revealing perspectives from clinical and research workflows},
  author={Verma, Himanshu and Mlynar, Jakub and Schaer, Roger and Reichenbach, Julien and Jreige, Mario and Prior, John and Ev{\'e}quoz, Florian and Depeursinge, Adrien},
  booktitle={Proceedings of the 2023 CHI conference on human factors in computing systems},
  pages={1--19},
  year={2023}
}

@inproceedings{gomez2025does,
  title={Why does Automation Adoption in Organizations Remain a Fallacy?: Scrutinizing Practitioners’ Imaginaries in an International Airport},
  author={Gomez-Beldarrain, Garoa and Verma, Himanshu and Kim, Euiyoung and Bozzon, Alessandro},
  booktitle={In CHI Conference on Human Factors in Computing Systems,(CHI’25), Association for Computing Machinery, New York, NY, USA},
  year={2025}
}

@article{smith2021interpretative,
  title={Interpretative phenomenological analysis: Theory, method and research},
  author={Smith, Jonathan A and Larkin, Michael and Flowers, Paul},
  year={2021},
  publisher={sAgE Publications ltd}
}

@article{guest2006many,
  title={How many interviews are enough? An experiment with data saturation and variability},
  author={Guest, Greg and Bunce, Arwen and Johnson, Laura},
  journal={Field methods},
  volume={18},
  number={1},
  pages={59--82},
  year={2006},
  publisher={Sage Publications Sage CA: Thousand Oaks, CA}
}

@article{mitchell2021automated,
  title={Automated vs. human health coaching: exploring participant and practitioner experiences},
  author={Mitchell, Elliot G and Maimone, Rosa and Cassells, Andrea and Tobin, Jonathan N and Davidson, Patricia and Smaldone, Arlene M and Mamykina, Lena},
  journal={Proceedings of the ACM on human-computer interaction},
  volume={5},
  number={CSCW1},
  pages={1--37},
  year={2021},
  publisher={ACM New York, NY, USA}
}

@book{Textor1995,
  author    = {Robert B. Textor},
  title     = {Ethnographic Futures Research: The Exploration of the Possible},
  year      = {1995},
  publisher = {Stanford University},
  address   = {Stanford, CA},
  note      = {Original method description of ethnographic futuring}
}

@inproceedings{EnglishLueck2021,
  author    = {J. A. English-Lueck and Sam Ladner and Liza Sherman},
  title     = {Little Dramas Everywhere: Using Ethnography to Anticipate the Future},
  booktitle = {Proceedings of EPIC 2021},
  year      = {2021},
  publisher = {EPIC},
  url       = {https://www.epicpeople.org/little-dramas-everywhere/}
}

@article{Elliott2017,
  author    = {Anthony Elliott and Brian McKelvey and Glenn Bowen},
  title     = {Marking Time in Ethnography: Uncovering Temporal Dispositions},
  journal   = {Time \& Society},
  volume    = {26},
  number    = {3},
  pages     = {314--336},
  year      = {2017},
  doi       = {10.1177/1466138116655360}
}

@article{Dawson2014,
  author    = {Patrick Dawson},
  title     = {Temporal Practices: Time and Ethnographic Research in Changing Organizations},
  journal   = {Journal of Organizational Ethnography},
  volume    = {3},
  number    = {2},
  pages     = {130--151},
  year      = {2014},
  doi       = {10.1108/JOE-05-2012-0025}
}

@article{baxter2011socio,
  title={Socio-technical systems: From design methods to systems engineering},
  author={Baxter, Gordon and Sommerville, Ian},
  journal={Interacting with computers},
  volume={23},
  number={1},
  pages={4--17},
  year={2011},
  publisher={OUP}
}

@book{trist1981evolution,
  title={The evolution of socio-technical systems},
  author={Trist, Eric L},
  volume={2},
  year={1981},
  publisher={Ontario Quality of Working Life Centre Toronto}
}

@article{winter2014beyond,
  title={Beyond the organizational ‘container’: Conceptualizing 21st century sociotechnical work},
  author={Winter, Susan and Berente, Nicholas and Howison, James and Butler, Brian},
  journal={Information and Organization},
  volume={24},
  number={4},
  pages={250--269},
  year={2014},
  publisher={Elsevier}
}

@article{bai2022training,
  title={Training a helpful and harmless assistant with reinforcement learning from human feedback},
  author={Bai, Yuntao and Jones, Andy and Ndousse, Kamal and Askell, Amanda and Chen, Anna and DasSarma, Nova and Drain, Dawn and Fort, Stanislav and Ganguli, Deep and Henighan, Tom and others},
  journal={arXiv preprint arXiv:2204.05862},
  year={2022}
}

@inproceedings{xiao2025might,
  title={" It Might be Technically Impressive, But It's Practically Useless to us": Motivations, Practices, Challenges, and Opportunities for Cross-Functional Collaboration around AI within the News Industry},
  author={Xiao, Qing and Fan, Xianzhe and Simon, Felix Marvin and Zhang, Bingbing and Eslami, Motahhare},
  booktitle={Proceedings of the 2025 CHI Conference on Human Factors in Computing Systems},
  pages={1--19},
  year={2025}
}

@article{zhan2024there,
  title={What is there to fear? Understanding multi-dimensional fear of AI from a technological affordance perspective},
  author={Zhan, Emily S and Molina, Mar{\'\i}a D and Rheu, Minjin and Peng, Wei},
  journal={International Journal of Human--Computer Interaction},
  volume={40},
  number={22},
  pages={7127--7144},
  year={2024},
  publisher={Taylor \& Francis}
}

@article{daugherty2019creating,
  title={Creating the symbiotic AI workforce of the future},
  author={Daugherty, Paul R and Wilson, H James},
  journal={MIT Sloan Management Review},
  volume={61},
  number={1},
  pages={1--4},
  year={2019},
  publisher={Massachusetts Institute of Technology, Cambridge, MA}
}

@article{abbass2019social,
  title={Social integration of artificial intelligence: functions, automation allocation logic and human-autonomy trust},
  author={Abbass, Hussein A},
  journal={Cognitive Computation},
  volume={11},
  number={2},
  pages={159--171},
  year={2019},
  publisher={Springer}
}

@article{mokyr2015history,
  title={The history of technological anxiety and the future of economic growth: Is this time different?},
  author={Mokyr, Joel and Vickers, Chris and Ziebarth, Nicolas L},
  journal={Journal of economic perspectives},
  volume={29},
  number={3},
  pages={31--50},
  year={2015},
  publisher={American Economic Association 2014 Broadway, Suite 305, Nashville, TN 37203-2418}
}

@article{fleming2019robots,
  title={Robots and organization studies: Why robots might not want to steal your job},
  author={Fleming, Peter},
  journal={Organization Studies},
  volume={40},
  number={1},
  pages={23--38},
  year={2019},
  publisher={Sage Publications Sage UK: London, England}
}

@inproceedings{weisz2025examining,
  title={Examining the use and impact of an ai code assistant on developer productivity and experience in the enterprise},
  author={Weisz, Justin D and Kumar, Shraddha Vijay and Muller, Michael and Browne, Karen-Ellen and Goldberg, Arielle and Heintze, Katrin Ellice and Bajpai, Shagun},
  booktitle={Proceedings of the Extended Abstracts of the CHI Conference on Human Factors in Computing Systems},
  pages={1--13},
  year={2025}
}

@inproceedings{wang2021brilliant,
  title={“Brilliant AI doctor” in rural clinics: Challenges in AI-powered clinical decision support system deployment},
  author={Wang, Dakuo and Wang, Liuping and Zhang, Zhan and Wang, Ding and Zhu, Haiyi and Gao, Yvonne and Fan, Xiangmin and Tian, Feng},
  booktitle={Proceedings of the 2021 CHI conference on human factors in computing systems},
  pages={1--18},
  year={2021}
}

@article{schmidt2002problem,
  title   = {The Problem with ‘Awareness’: Introductory Remarks on Awareness in CSCW},
  author  = {Schmidt, Kjeld},
  journal = {Computer Supported Cooperative Work (CSCW)},
  volume  = {11},
  number  = {3--4},
  pages   = {285--298},
  year    = {2002},
  publisher = {Springer}
}

@incollection{suchman1995making,
  title     = {Making Work Visible},
  author    = {Suchman, Lucy},
  booktitle = {Communications of the ACM},
  volume    = {38},
  number    = {9},
  pages     = {56--64},
  year      = {1995},
  publisher = {ACM}
}

@article{sadeghian2025workai,
  title={WorkAI: A Toolkit for the Design of AI-driven Future of Work},
  author={Sadeghian, Shadan and Bareikyte, Migle and Burkhardt, Marcus and Hassenzahl, Marc},
  journal={Proceedings of the ACM on Human-Computer Interaction},
  volume={9},
  number={7},
  pages={1--27},
  year={2025},
  publisher={ACM New York, NY, USA}
}

\appendix

\end{document}